\documentclass[preprint, preprintnumbers,amsmath,amssymb,prd,superscriptaddress,nofootinbib]{revtex4}

\usepackage{graphicx}
\usepackage{bm}
\usepackage{color}
\usepackage{multirow}

\begin{document}

\newcommand{\Osaka}{
  Department of Physics, The University of Osaka,
  Toyonaka 560-0043, Japan
}
\newcommand{\Edinburgh}{
  School of Physics and Astronomy,
  The University of Edinburgh,
  Edinburgh EH9 3JZ, United Kingdom 
}
\newcommand{\KEK}{
  High Energy Accelerator Research Organization (KEK), 
  Tsukuba 305-0801, Japan
}
\newcommand{\Sokendai}{
  School of High Energy Accelerator Science,
  The Graduate University for Advanced Studies (Sokendai), 
  Tsukuba 305-0801, Japan
}
\newcommand{\RIKENCCS}{
  RIKEN Center for Computational Science,
  7-1-26 Minatojima-minami-machi, Chuo-ku, Kobe, Hyogo 650-0047, Japan
}
\newcommand{\JAEA}{
  Advanced Science Research Center,
  Japan Atomic Energy Agency (JAEA),
  Tokai 319-1195, Japan
}
\preprint{OU-HET-1250}

\title{
  Symmetry of screening masses of mesons in two-flavor lattice QCD at high temperatures
}
\author{Y.~Aoki}
\affiliation{\RIKENCCS}
\author{H.~Fukaya}
\affiliation{\Osaka}
\author{S.~Hashimoto}
\affiliation{\KEK}
\affiliation{\Sokendai}
\author{I.~Kanamori}
\affiliation{\RIKENCCS}
\author{Y.~Nakamura}
\affiliation{\RIKENCCS}
\author{C.~Rohrhofer}
\affiliation{\Osaka}
\author{K.~Suzuki}
\affiliation{\JAEA}
\author{D.~Ward}
\affiliation{\Osaka}

\collaboration{JLQCD collaboration}
\noaffiliation

\begin{abstract}
  We investigate spatial two-point correlation functions
  of mesonic operators
  in two-flavor lattice QCD at high temperatures.
  The simulated temperatures cover 
  the range $T \in [147, 330]$ MeV, where
  the critical temperature is estimated around 165 MeV.
  To ensure a good control of the chiral symmetry
  we employ the M\"obius domain-wall fermion action for
  two degenerate flavors of quarks.
  With a lattice cut off $a^{-1}\sim 2.6$ GeV, 
  the residual mass is reduced to  0.14 MeV.
  With the energy spectrum obtained from the screening mass at incremental values of the 
  temperature range, we examine the $SU(2)_L\times SU(2)_R$
  chiral symmetry, the anomalous axial $U(1)$ as well as an 
  enhanced symmetry which exchanges the spin degrees of freedom.
  We also study how the data approaches the perturbative prediction 
 given by twice the Matsubara frequency of free quarks.
\end{abstract}
\maketitle

\section{Introduction}

The chiral phase transition of quantum chromodynamics (QCD), 
which occurred in the early universe,
gives a crucial input for understanding how
hadrons formed and acquired their $O(1)$ GeV masses we observe currently.
While it is naturally assumed that the lightest two quarks flavors
play a crucial role in the phase transition, 
the details of the critical phenomena depend on 
the symmetry of the up and down quarks above the critical temperature $T_c$ \cite{Pisarski:1983ms}.

In the massless limit of the up and down quarks, 
it is natural to assume that the standard $SU(2)_L\times SU(2)_R$ symmetry
is recovered at some critical temperature in the range 150--200 MeV
and no other symmetry exists.
However, the axial $U(1)_A$, which is broken by anomaly, 
may effectively appear
if nontrivial topological excitations of gluons 
are strongly suppressed above $T_c$.
When such topological effects are described by instantons
\cite{Gross:1980br, Diakonov:1984vw,  Morris:1984zi, Schafer:1996wv,Ringwald:1999ze},
which is valid for pure Yang-Mills, QCD with heavy quarks, and QCD with a large number of color degrees of freedom, strong suppression of the quantum anomaly 
is not expected and the axial $U(1)_A$ is likely to remain broken at $T\sim T_c$.
However, with the light dynamical quarks, the typical size
of the topological excitation is described by the pion physics
whose correlation length becomes larger than $1/T$, and this description of the topological excitations by instantons is no longer valid.
The fate of the axial $U(1)_A$ anomaly at high $T$ has been discussed
in both theoretical works 
\cite{Cohen:1996ng,Cohen:1997hz,Aoki:2012yj,Pelissetto:2013hqa,Kanazawa:2014cua,Sato:2014axa,Nakayama:2016jhq}
as well as numerical works \cite{Cossu:2013uua,Buchoff:2013nra,Dick:2015twa,Brandt:2016daq,Ishikawa:2017nwl,Bazavov:2019www,Kaczmarek:2023bxb}. 

\if
Another interesting possibility is that at high $T$, QCD may in fact have a larger symmetry. This symmetry is related to the anti-periodic boundary condition of quarks imposed in the temporal direction.
At very high temperature, the mesons would have large masses proportional to 
the temperature $T$, due to the Matsubara mass. 
\if
#1 - Reviewer point
L&V analysis of 2pi T+g^2T has been checked, MWDF is an important 
Add a section to describe the fact that our lattice sapcing is fine
Since the symmetry is exact we do not have to worry about the discretization effect
#2 - Reviewer Point 
Review and possibly omit 0.00375 S channel based on health of the correlator
Clarify mass or mention pseudocritical temp shift based on mass
Mass independence of restored symmetry for temps above Tpc

\fi
As a consequence, a nontrivial symmetry 
may appear, analogous to the heavy quark symmetry \cite{Neubert:1993mb, Brambilla:2004jw}
at $T=0$.
In fact, in the perturbative evaluation in \cite{Laine:2003bd}
the masses of all the mesons of different spins converge to $2\pi T$ at lowest order.
Note that this is not a symmetry of the original QCD Lagrangian
but is an ``emergent'' symmetry in the large $T$ limit.
Although this is only an approximate
symmetry within the perturbative framework in the large $T$ limit,
nonperturbative gluonic effects may enhance this symmetry
even at $T\sim 2T_c$, this symmetry is called the chiral-spin symmetry
\cite{Glozman:2014mka, Glozman:2016swy, Glozman:2017dfd, Lang:2018vuu, Catillo:2021rrq, Glozman:2022lda}.
\fi

Another interesting possibility is that at high $T$, QCD may in fact have a larger symmetry. This symmetry is related to the anti-periodic boundary condition of quarks imposed in the temporal direction.
At very high temperature, the mesons would have large masses proportional to 
the temperature $T$, due to the Matsubara mass. 
In fact, in the perturbative evaluation in \cite{Laine:2003bd}
the masses of all the mesons of different spins converge to $2\pi T$ at lowest order. From this point of view, we could consider the light quarks which form the meson to become sufficiently ``heavy'' at high temperatures leading to a spin independent symmetry. Such a symmetry is analogous to the heavy quark symmetry \cite{Neubert:1993mb, Brambilla:2004jw} in $T=0$ QCD.  Note that this is not a symmetry of the original QCD Lagrangian
but is an ``emergent'' symmetry in the large $T$ limit.
Although this is only an approximate
symmetry within the perturbative framework in the large $T$ limit,
nonperturbative gluonic effects may enhance this symmetry
even at $T\sim 2T_c$, this symmetry is called the chiral-spin symmetry
\cite{Glozman:2014mka, Glozman:2016swy, Glozman:2017dfd, Lang:2018vuu, Catillo:2021rrq, Glozman:2022lda}.

Our goal in this work is to simulate QCD with chiral fermions and 
to investigate these symmetries at high temperatures,
including how the meson masses approach the perturbative prediction $2\pi T$\footnote{
In \cite{DallaBrida:2021ddx}, they simulated QCD at very high temperatures $T\ge 1$ GeV
and found that the meson masses approach $2\pi T$ from above,
which is consistent with \cite{Laine:2003bd} although the channel-dependence is still visible.
}.
Lattice QCD simulations that have previously studied this issue, 
however, used fermion formulations that explicitly
broke the $SU(2)_L\times SU(2)_R$ symmetry.
For the Wilson fermion, the symmetry is broken down to a vectorlike $SU(2)_V$
and for the staggered fermion, only the $U(1)_{V'}\times U(1)_{A'}$ subgroup
(which are different from the standard quark number and anomalous axial $U(1)_A$ symmetries) remains.
In particular, we reported in \cite{Cossu:2015kfa, Tomiya:2016jwr, Aoki:2020noz}
that the lattice artifacts in the probes of 
axial $U(1)_A$ anomaly are enhanced at high temperatures
and it is difficult to disentangle the physical effect 
from the discretization errors.

In this work, we employ the domain-wall fermion formalism
\cite{Kaplan:1992bt,Shamir:1993zy,Furman:1994ky}
with an improvement proposed in Refs.\cite{Brower:2005qw,Brower:2012vk}.
With this M\"obius domain-wall fermion, which is
a good approximation of the overlap fermion \cite{Neuberger:1997fp},
we achieve a theoretically clean control of the symmetry of QCD. 
With a fixed lattice spacing $a\sim 0.075$ fm, 
the residual mass \cite{Tomiya:2016jwr}, which is an indicator of the 
explicit symmetry violation is reduced to 0.14(6) MeV.
In addition, we set the quark masses so that the 
lightest bare quark mass $ma=0.001$ is less than the physical mass value
of the up and down quarks.
The simulated temperatures span a range 
$T \in [147, 330]$ MeV, which corresponds to $[0.9, 2] T_c$,
where the $T_c$ is estimated from the peak of the chiral susceptibility \cite{Aoki:2024uvl}.
The main runs are carried out with a fixed lattice size $L=32$
except for the lowest two temperatures, where the pions 
behave as the pseudo-Nambu-Goldstone bosons;
in this case we simulate two additional volumes,
$L=40$ and $48$, to control the finite volume systematics. 
While this study is carried out for a fixed lattice spacing, our lattice spacing 
$a\sim 0.075$ fm is at the same level of the finest lattice in the literature.
In addition to this, our use of the M\"obius domain wall fermion introduces an automatic $\mathcal{O}(a)$ improvement over dicretization effects; while these cannot be estimated in this study, future work will be to make an extrapolation to the continuum limit.

On each ensemble of O(1000-10000) trajectories,
we measure the two-point mesonic correlation functions
in 6 different iso-triplet channels every 50 trajectories. 
With the screening masses obtained from fitting the correlators, 
we examine the symmetry at each simulated temperature and quark mass.
In our main analysis of the mesonic correlators,
the standard $SU(2)_L\times SU(2)_R$ chiral symmetry
is examined by the difference between 
the vector $V$ and axial vector $A$ correlators.
The tensor $T_t$ and $X_t$ channels (definitions will be given below), 
rather than scalar $S$(noisy) and pseudoscalar $PS$ channels,
are useful to estimate how much the axial $U(1)_A$ anomaly effect remains.
In lattice QCD, the isospin triplet $S$ channel has been difficult to precisely measure. As a possible reason, it was pointed out in \cite{Mathur:2006bs} that the isospin triplet $S$ channel may couple to many composite meson states. Therefore, study of the $U(1)_A$ as well as any other underlying quark symmetries using the $S$ channel should be considered alongside an additional probe.
Because of the Matsubara mass, 
the system may exhibit the chiral-spin $SU(2)_{CS}$ symmetry mentioned above.
In our previous work at higher temperatures \cite{Rohrhofer:2017grg, Rohrhofer:2019qwq, Rohrhofer:2019qal}
we observed the emergence of this $SU(2)_{CS}$ symmetry from
the correlator ratios among $V$, $A$, $T_t$ and $X_t$.
A similar observation simulating quarks with a good chiral symmetry
was reported in Refs.~\cite{Chiu:2023hnm,Chiu:2024jyz,Chiu:2024bqx}.
In this work, we revisit this enlarged symmetry group,
computing the screening masses and simulating lower temperatures
down to $T\sim 147$ MeV, which is $\sim 0.9 T_c$.

The rest of the paper is organized as follows.
In Sec.~\ref{sec:corr}, we disucuss the general functional form of the mesonic correlators 
and the possible pattern of the symmetry at high temperatures.
In particular, we address the enhancement of symmetry compared to the original QCD 
using the effective theory of heavy quarks due to the Matsubara mass.
Our lattice setups as well as our methodology for  
how to extract the screening mass are addressed in Sec.~\ref{sec:lattice}.
Numerical results for our lattices are given in Sec.~\ref{sec:results}.
And finally, we give a summary and discuss our numerical results in Sec.~\ref{sec:summary}.
Some results in this work were already given in \cite{Ward:2024wze} and
preliminary results for $N_f=2+1$ simulations were presented in \cite{Ward:2024tdm}.
We also refer the readers to Refs.~\cite{Aoki:2021qws, JLQCD:2024xey, Goswami:2024kcq}
for the related works from the JLQCD finite $T$ project.
\section{Mesonic two-point functions and symmetries at high temperatures}
\label{sec:corr}



In the limit of massless up and down quarks,
the QCD action is invariant under the standard chiral transformation
of the group $SU(2)_L\times SU(2)_R$, as well as
that of the axial $U(1)_A$.
The axial $U(1)_A$ symmetry is in general broken by quantum anomaly,
but its magnitude at finite temperatures is nontrivial since it is tightly related
to the topological excitations of gluons,
which may be strongly suppressed at high temperatures.

In this work, we consider  the mesonic two-point function in the
$z$ direction whose continuum limit is given by
\begin{equation}
  C_\Gamma(z) = \int_{-\infty}^\infty dx \int_{-\infty}^\infty dy \int_{0}^\beta dt
  \left\langle O_\Gamma(x,y,z,t)O_\Gamma^\dagger(0,0,0,0)\right\rangle,
\end{equation}
where the quark bilinear operator $O_\Gamma=\bar{q}\Gamma T^a q$
is taken to be a flavor-triplet (the isospin generator is denoted by $T^a$).
For  $\Gamma$ we choose $\gamma_5(PS), \mathbb{I}(S), \gamma_{1,2}(V), 
\gamma_5\gamma_{1,2}(A), \gamma_4\gamma_3(T_t)$
and $\gamma_5\gamma_4\gamma_3(X_t)$.
The list of operators we measure and the symmetries connecting them 
are summarized in Tab.~\ref{tab:sympat}.

Under the standard chiral $SU(2)_L\times SU(2)_R$ rotation:
\begin{align}
  q \to \exp(i\alpha_aT^a +i\beta_aT^a\gamma_5  ) q,\;\;\;
  \bar{q} \to \bar{q}\exp(-i\alpha_aT^a +i\beta_aT^a\gamma_5),
\end{align}
only $V$ and $A$ correlators mix 
while all the other channels remain unchanged up to isospin exchanges.
Therefore, the $SU(2)_L\times SU(2)_R$ symmetry breaking can be detected by 
the difference between the $V$ and $A$ correlators.

For the axial $U(1)_A$ transformation,
\begin{align}
  q \to \exp(i\epsilon \gamma_5  ) q,\;\;\;
  \bar{q} \to \bar{q}\exp(i\epsilon\gamma_5),
\end{align}
$S$ and $PS$ mix as well as the $T_t$ and $X_t$ channels,
while $A$ and $V$ correlators are invariant.
Thus, the $S$--$PS$ and $T_t$--$X_t$ pairs
are the probes of the axial $U(1)_A$ breaking.
As shown later, the $S$ correlator is numerically noisy, and 
$T_t$ and $X_t$ are more useful to examine the axial $U(1)_A$ anomaly.

\begin{table}[h]
 \centering
 \begin{tabular}{c c c c c}
  \hline
  $\Gamma$ & Reference Name & Abbr. & \multicolumn{2}{c}{Symmetry Correspondences}\\
  \hline
  $\mathbb{I}$ & Scalar & S & \multirow{2}{*}{$\bigg\}U(1)_A$}\\
  $\gamma_5$ & Psuedo Scalar & PS\\
  $\gamma_{1},\gamma_2$ & Vector & $V$ & \multirow{2}{*}{$\bigg\}SU(2)_L \times SU(2)_R$} \\
  $\gamma_{1}\gamma_5,\gamma_2\gamma_5$ & Axial Vector & $A$ & & \multirow{3}{*}{{\large $\Bigg\}$} $SU(2)_{CS}$ }\\
  $\gamma_4\gamma_3$ & Tensor &  $T_t$ & \multirow{2}{*}{$\bigg\}U(1)_A$} \\
  $\gamma_4\gamma_3\gamma_5$ & Axial Tensor &  $X_t$\\
  \hline
 \end{tabular}
\caption{List of quark bilinear operators we compute the two-point correlation functions and the symmetries connecting them. \label{tab:sympat}}
\end{table}

In addition to the above manifest symmetries in the QCD Lagrangian,
we may expect emergence of larger symmetries
due to anti-periodic boundary condition on the quarks
in the temporal direction.
This enhancement of symmetry is analogus to
the heavy quark symmetry \cite{Neubert:1993mb, Brambilla:2004jw}, 
which (approximately) appears in the
effective theory of bottom or charm quarks.

In order to see this enhanced symmetry, let us consider a
free quark propagator in the positive $z$ direction:
\begin{align}
 \langle q(z)\bar{q}(0)\rangle(p_1,p_2) &= T\sum_{p_0} \int \frac{dp_3}{(2\pi)}
\frac{e^{ip_3 z}}{i\gamma_0p_0 + i\gamma_3 p_3 +i \gamma_1 p_1+i \gamma_2 p_2 +m}
\nonumber\\
&=  T\sum_{p_0} \int \frac{dp_3}{(2\pi)}
\frac{-(i\gamma_0p_0 + i\gamma_3 p_3 +i \gamma_1 p_1+i \gamma_2 p_2 -m)e^{ip_3 z}}{p_0^2+p_3^2+p_2^2+p_1^2+m^2}
\nonumber\\
&=  T\sum_{p_0} 
\frac{-(i\gamma_0p_0 -\gamma_3 E(p_0,p_1,p_2)  +i \gamma_1 p_1+i \gamma_2 p_2 -m)e^{-E(p_0,p_1,p_2)z}}{2E(p_0,p_1,p_2)}
\end{align}
where $E(p_0,p_1,p_2)=\sqrt{p_0^2+p_1^2+p_2^2+m^2}$ and the other components $p_1,p_2$ are fixed.
Here the Matsubara mass $p_0=(2n+1)\pi T$ takes discrete values due to 
the antiperiodicity in the imaginary temporal direction.
In the large $T\gg m,p_1,p_2$ limit, 
the lowest $p_0=\pm \pi T$ will dominate the signal
and the correlator becomes
\begin{align}
\label{eq:qcorrz}
 \langle q(z)\bar{q}(0)\rangle(p_1,p_2) 
&= T\left[\gamma_3
\frac{1+i\text{sgn}(p_0)\gamma_0\gamma_3}{2}e^{-\pi Tz}+O(1/T)\right].
\end{align}

In the same way, the correlator in the temporal direction with 
a fixed $p_1,p_2,p_3$ in the large Matsubara mass limit is expressed as
\begin{align}
 \langle q(t)\bar{q}(0)\rangle(p_1,p_2,p_3) 
&= -iT\left[\sum_{p_0} \frac{\gamma_0 p_0}{p_0^2}\exp(ip_0 t) +O(1/T)\right]. 
\end{align}

These propagators are invariant under the following transformation:
\begin{align}
  q &\to \exp(i\Sigma^i \epsilon_i) q,
  \nonumber\\
  \bar{q} &\to \bar{q}\gamma_0\exp(-i\Sigma^i \epsilon_i)\gamma_0,
\end{align}
where $\Sigma^i$ has three components,
\begin{equation}
\Sigma^i=(\gamma_k, -i\gamma_5\gamma_k, \gamma_5),\;\;\; k=1,2.
\end{equation}
We do not distinguish $k=1$ and $k=2$, since 
they are identical under the rotational symmetry along the $z$ axis.
The generators $\Sigma^i$ form an $SU(2)$ algebra, which is identical to the chiral-spin $SU(2)_{CS}$ symmetry suggested in the literature
\cite{Glozman:2014mka, Glozman:2016swy, Glozman:2017dfd, Lang:2018vuu, Catillo:2021rrq, Glozman:2022lda}\footnote{
In the hadron correlators, the relative momentum part in $p_1$ and $p_2$ must be integrated out
but that ends up with a typical scale $1/z$ and the correction is $1/zT$ 
as clearly shown in \cite{Rohrhofer:2019qwq}.
}.

Note that any correction from gauge field potential
only appears in the next-to-leading $1/T$ contribution.
Thus, the chiral spin $SU(2)_{CS}$ symmetry emerges 
at sufficiently high temperatures.
With this $SU(2)_{CS}$ symmetry, the mesonic operators
$(A_1, T_t, X_t)$ form a nontrivial triplet.
These operators have different spins in the original four dimensions,
but they are all in the spin-one representation
of $SO(3)$ symmetry along the $z$ axis.
One can consider another set $(V_2, PS, S)$ 
but the operators transform differently under the $SO(3)$ rotation
and therefore, the mass degeneracy would not be that good.

From (\ref{eq:qcorrz}), it is obvious that
in the free quark limit the meson screening mass is $2\pi T$. 
In fact, in the perturbative analysis \cite{Laine:2003bd}
the screening mass is obtained as
\[
 M_\Gamma = 2\pi T + cg^2T, 
\]
with a numerical constant $c$ coming from the one-loop correction
where $g$ is the gauge coupling.
It is remarkable that the result does not depend on $\Gamma$.
This result is analogous to the heavy quark symmetry \cite{Neubert:1993mb, Brambilla:2004jw}
which is an approximate
symmetry of the bottom or charm quarks at $T=0$.
As the spin and orbital angular momentum dependent interactions are 
proportional to the inverse of the heavy quark mass,
the spectrum of the quarkonia and heavy-light mesons are insensitive to spin.
The same may explain here that with a large Matsubara mass proportional to $T$,
the operators $(V_1, PS, S)$ with different spin form an $SU(2)_{CS}$ triplet\footnote{
The $SO(3)$ rotational symmetry is broken also among $(A_1, T_t, X_t)$
at very high temperatures, since the temporal direction shrinks as $1/T\to 0$.
However, its effect is sub-leading.
}.
In Ref.~\cite{Chiu:2023hnm}, it was suggested that
other combinations of tensors $\gamma_1\gamma_3(T_k)$
and $\gamma_1\gamma_3\gamma_5(X_k)$ form a quartet with $A_4$ and $V_4$.
Inclusion of this quartet into our analysis is of interest to future work; however, in this paper we will focus on the triplet.

It is an important numerical subject to quantify 
the emergent $SU(2)_{CS}$ symmetry at intermediate temperatures 
$T\sim T_c$ and see if it has any impact on the chiral phase transition.
It is also interesting to investigate if an additional transition exists at $T>T_c$, triggered by this emergent symmetry.
These three symmetries, accessible through the differences between mesonic screening masses from the pairs of the triplet channels\footnote{
In our previous study \cite{Rohrhofer:2019qal},
we also studied the ratio of differences
of correlators $|C_A(z)-C_T(z)|/|C_A(z)-C_S(z)|$. 
However, we find that the denominator approaches to zero
at high temperatures (presumably due to the large Matsubara mass and the analogy to 
``heavy quark symmetry'' we discussed in Sec.~\ref{sec:corr}.) 
and the ratio is not a good indicator of the $SU(2)_{CS}$ symmetry. 
In this work, we focus on the screening mass differences
and present the values in the physical unit (MeV)
so that one can easily compare with the simulated temperature for each set of data. 
}, constitute our motivation to simulate QCD at high temperatures in the range 
$[0.9,2]T_c$, for a range of quark masses covering the physical point as a method to understand the symmetry structure of QCD at high temperatures.  

\section{Lattice setups}
\label{sec:lattice}

For gauge configuration generation, we perform hybrid Monte Carlo
simulations\footnote{See Refs.~\cite{Cossu:2013ola,Boyle:2015tjk,Ueda:2014rya,Amagasa:2015zwb} 
for details of our simulation codes.} 
with the tree-level improved
Symanzik gauge action \cite{Luscher:1985zq}
and the M\"obius domain-wall fermion action \cite{Brower:2005qw,Brower:2012vk}.
See Ref.~\cite{Tomiya:2016jwr} for details of our simulations.
In this work, we focus on the gauge configurations
at a fixed lattice spacing $0.075$ fm generated with the
bare coupling constant $\beta=4.30$.
A portion of the same gauge link ensembles are shared with
our previous studies~\cite{Aoki:2021qws, Aoki:2020noz}, in addition to these
we also use an updated and newly generated series of ensembles on larger lattices
at lower temperatures.
The simulation parameters are summarized in Tab.~\ref{tab:parameters-results}.

For the scale setting, we perform the Wilson flow
and use the reference flow time $t_0=(0.1539\;\mbox{fm})^2$
presented in \cite{Sommer:2014mea} to determine the lattice cut-off $1/a=2.643$ GeV.
Our main lattice size $L=32$ corresponds to 2.37 fm, while we also simulate
$L=40$ and 48 lattices for the lowest two simulated temperatures
to study the finite size systematics near the critical temperature.

We vary the temperature by changing the temporal size of the lattice
from 8 to 18, which corresponds the range 147 MeV $\le T \le$ 330 MeV.
The critical temperature at the physical point is 
estimated to be $T_c \sim 165(3)$ MeV from the peak position of 
the disconnected chiral susceptibility \cite{Aoki:2024uvl} 
where the error is statistical only.

We take 4--6 different quark masses depending on the temperatures.
The lightest bare sea quark mass $m=0.001$ corresponds to 2.6 MeV, which is 
estimated to be 71\% of the physical quark mass $m_\text{phys}=0.0014(2)$.
In this study, we always take the valence and sea quark masses equal.

For the link variables in the Dirac operator,
we perform the stout smearing \cite{Morningstar:2003gk}
three times with the smoothing parameter $\rho=0.1$.
The size of the extra dimension for the M\"obius domain-wall fermion action is set to $L_s=16$.
With this treatment, the residual mass or the chiral symmetry breaking 
of the lattice Dirac operator is suppressed to $0.14(6)$ MeV \cite{Tomiya:2016jwr}.

For the correlator measurement, we use the four-dimensional effective Dirac operator,
\begin{equation}
 D_\text{DW}^\text{4D}(m)=\frac{1+m}{2}+\frac{1-m}{2}\gamma_5\tanh(L_s\tanh^{-1}(H_M)),
\end{equation}
where the M\"obius kernel $H_M$ is 
\begin{equation}
 H_M=\gamma_5\frac{2D_W}{2+D_W},
\end{equation}
and $D_W$ is the standard Wilson-Dirac operator with a negative mass $M=-1$.
Here and in the following, we set $a=1$ unless explicitly specified.

For each ensemble, 40-700 measurements were performed per 50 trajectories 
and our numerical data, after binning between 2-10 samples, did 
not show any significant autocorrelations.
Using the rotational symmetry of the z-directional correlators, we applied averaging over three different spatial directions. For our sources, we did not apply any additional source smearing and used point-like source and sink operators.

\section{Numerical results}
\label{sec:results}
To obtain the screening masses from the correlation functions the effective mass is extracted from data by applying Newton's method to the cosh ansatz form of the spatial correlator. We then perform an uncorrelated least square fit by finding a suitable range for which the effective mass does not exceed $2\sigma$ from the error band. Regions where the error became 100\% are omitted from the fits where possible. 
When the correlators are stable, the channels connected by symmetry transformations e.g. $V-A$, $PS-S$ and $X-T$ are fitted to the same range. We also confirm that the screening mass differences do not depend on the fitting range very much and the systematics are well controlled compared to the statistical uncertainties.

We find that the $S$ channel correlators are quite unstable, reflecting exceptionally large contribution from
a few configurations. 
We decide to discard the $S$ data when 1)the correlator goes negative 
or 2)there is no stable plateau in the fitting range of the symmetry partner $PS$.

\begin{figure*}
  \centering
  \includegraphics[width=0.4\columnwidth]{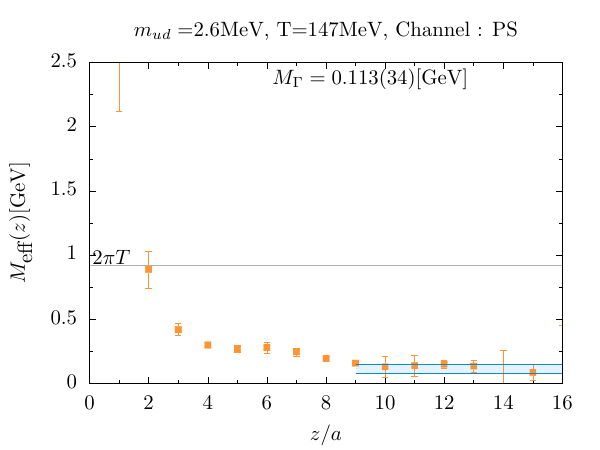}
\includegraphics[width=0.4\columnwidth]{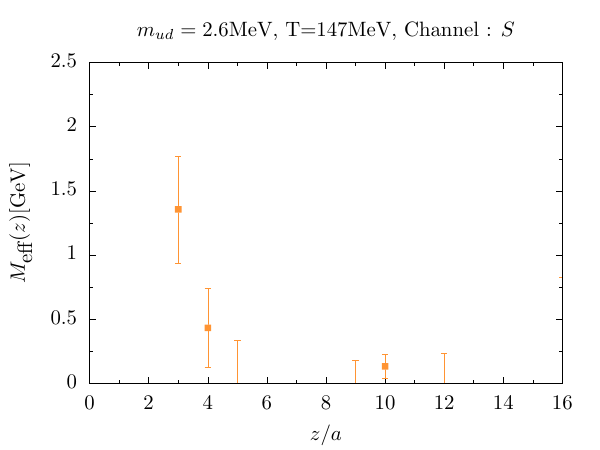}
\includegraphics[width=0.4\columnwidth]{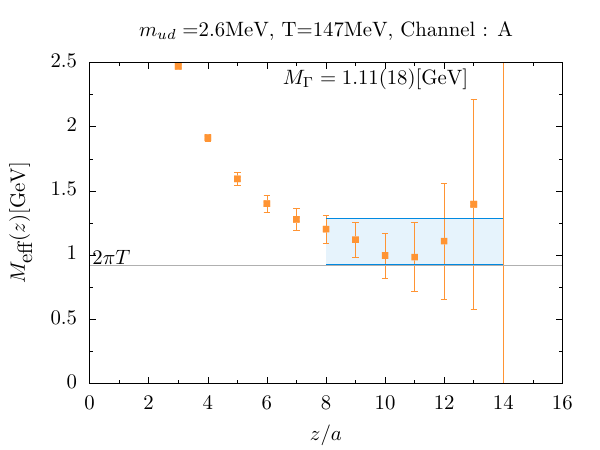}
\includegraphics[width=0.4\columnwidth]{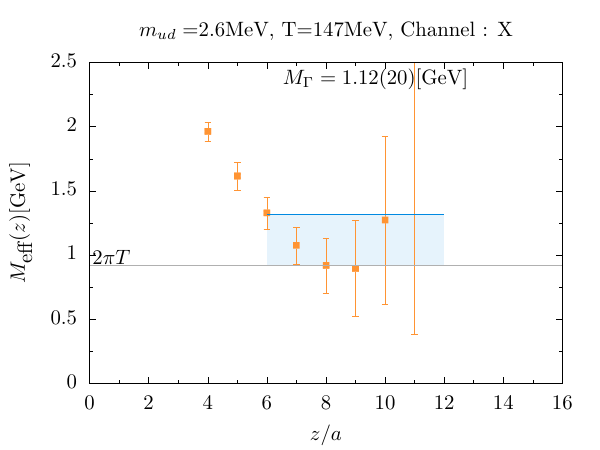}
  \caption{
    Effective screening mass plots at $T=147$ MeV.
 The data for $PS, S, A, X_t$ channels at the lightest quark mass $m=0.001$ are presented.
  }
  \label{fig:mesoneffmassT147}
\end{figure*}

\begin{figure*}
  \centering
  \includegraphics[width=0.4\columnwidth]{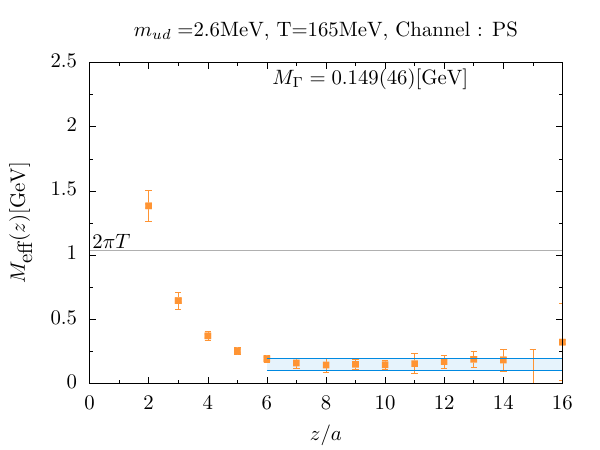}
\includegraphics[width=0.4\columnwidth]{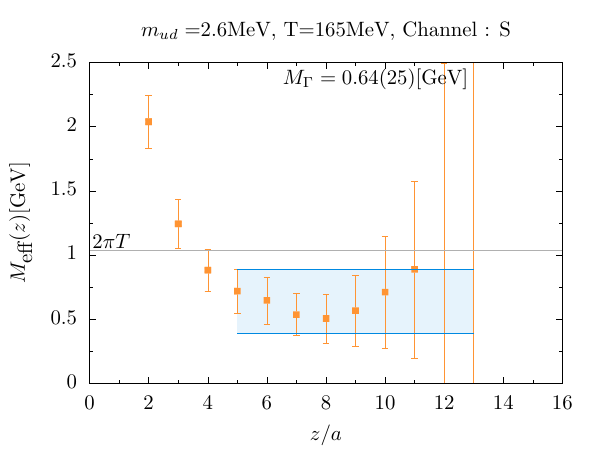}
\includegraphics[width=0.4\columnwidth]{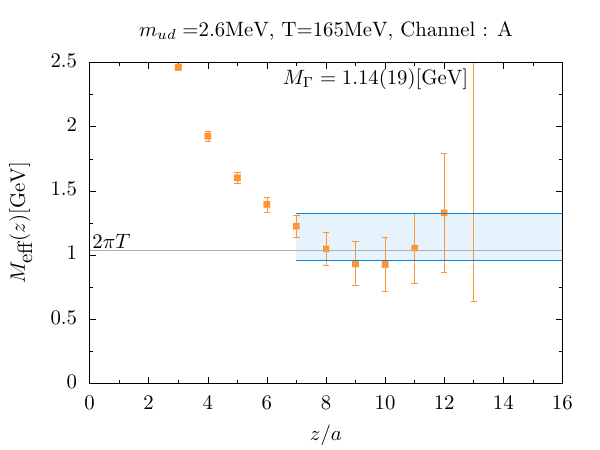}
\includegraphics[width=0.4\columnwidth]{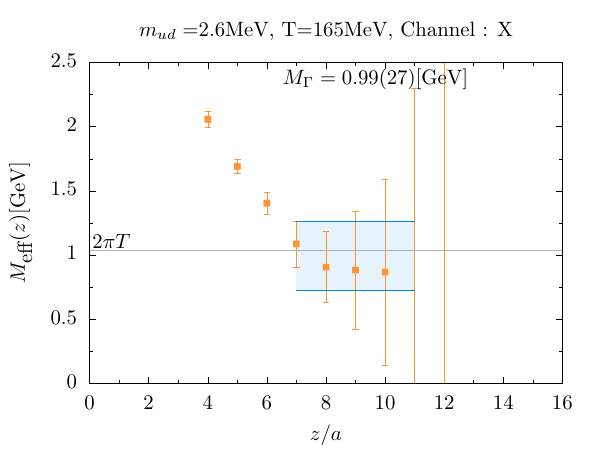}
  \caption{
 The same plots as Fig.~\ref{fig:mesoneffmassT147} but at $T=165$ MeV.
  }
  \label{fig:mesoneffmassT165}
\end{figure*}

\begin{figure*}
  \centering
  \includegraphics[width=0.4\columnwidth]{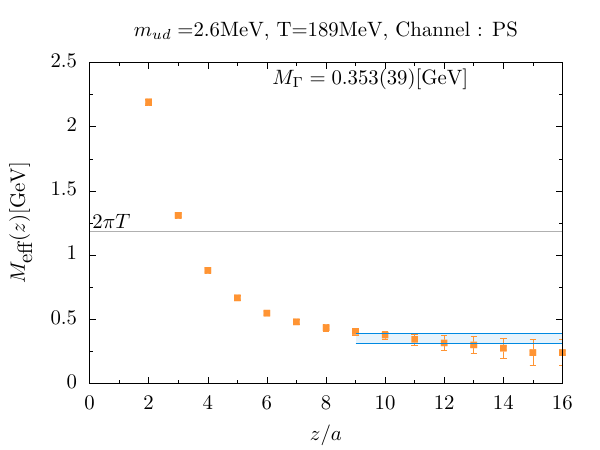}
\includegraphics[width=0.4\columnwidth]{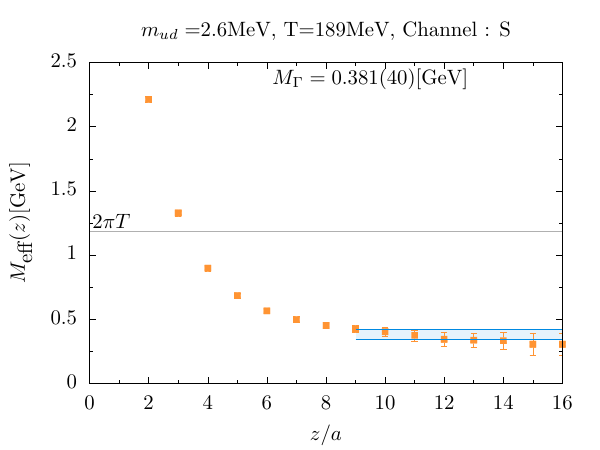}
\includegraphics[width=0.4\columnwidth]{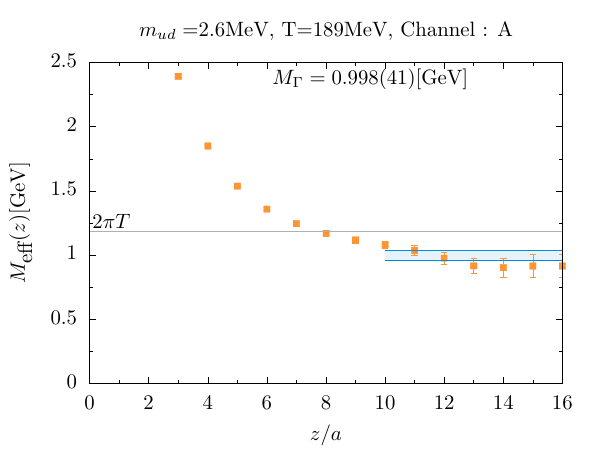}
\includegraphics[width=0.4\columnwidth]{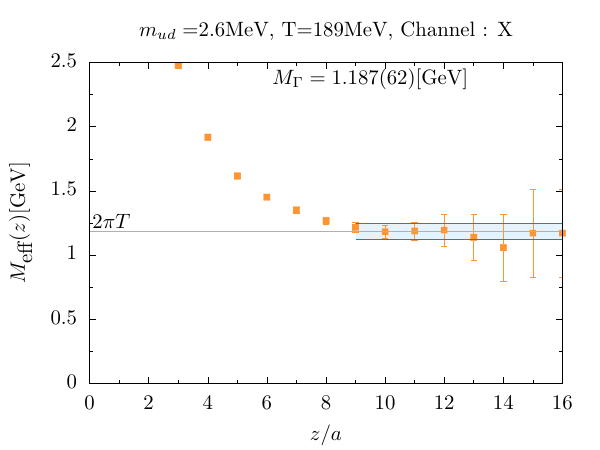}
  \caption{
 The same plots as Fig.~\ref{fig:mesoneffmassT147} but at $T=190$ MeV.
  }
  \label{fig:mesoneffmassT190}
\end{figure*}

Figures~\ref{fig:mesoneffmassT147}--\ref{fig:mesoneffmassT190}
present the typical effective mass plots with the $\cosh$ fitting ansatz
 at the lightest quark mass $m=0.001$ at the lowest three simulated temperatures
$T=$ 147, 165 and 190 MeV\footnote{We also present the corresponding raw correlators in the Appendix.}.
It is not difficult to find a reasonable plateau, in contrast to the study with staggered fermions in Ref.~\cite{Bazavov:2019www},
where a sizable contamination from the excited states was reported.
A possible reason for such a difference is the fact that
the M\"obius domain-wall fermion is free from the unphysical 
taste degrees of freedom as well as unwanted operator mixing
among different multiplets of the $SU(2)_L \times SU(2)_R$ flavor group.
From the data in the fitting range, which is shown by the shadow in these plots,
we determine the meson screening masses for the different channels.
The obtained values and their differences among various symmetry 
multiplets are summerized in Tab.~\ref{tab:parameters-results} and \ref{tab:sym-results}.

\begin{figure*}
  \centering
  \includegraphics[width=0.7\columnwidth]{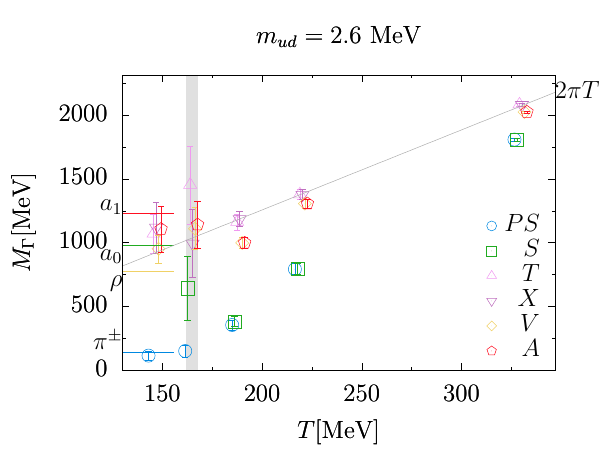}
  \caption{
$T$ dependence of the screening masses for various meson channels at the lightest simulated mass $m=0.001$($\sim $2.6 MeV).
The grey line shows $2\pi T$, which is twice of the Matsubara mass.
For a reference, we put the experimental results of $\pi^\pm$, $\rho$, $a_0$ and $a_1$ mesons
at zero temperature, indicated by bars at the vertical axis.
The vertical grey band indicates our estimate for the critical temperature $T_c=165(3)$ MeV from the 
chiral susceptibility. Note that the busy data points are slightly shifted horizontally. 
  }
  \label{fig:mscreenT}
\end{figure*}

In Fig.~\ref{fig:mscreenT},
we plot the temperature dependence of the obtained meson screening masses
at our lightest mass $m=0.001$ ($\sim$ 2.6 MeV).
As disscused in Sec.~\ref{sec:corr}, 
the data at high $T$ for all channels approach $2\pi T$,  
which is twice of the lowest mode Matsubara mass, shown by the grey line.
At lower temperatures around or below the critical temperature 
$T_c\sim  165$(3) MeV, 
shown by the vertical grey band,
there is a drastic change in the flucutation of the masses.
The scalar mass, in particular, 
shows inflated errorbars at the lower temperatures, while quickly converging to a less noisy result, close to the $PS$ channel data at higher temperatures.
For reference, we put the experimental results of $\pi^\pm$, $\rho$, $a_0$ and $a_1$ mesons
at zero temperature, indicated by colored bars corresponding to the associated channels on the vertical axis.
It is interesting to note that the lowest temperature, $T=147$ MeV 
data are already quite close to the experimental values at $T=0$;
this may indicate that the chiral symmetry breaking by the quark condensate 
is already strong enough to form hadrons.
In the $T=0$ limit, the $T_t$ and $X_t$
two-point functions transform in the same way as
$V$ and $A$ correlators, respectively, 
and share the same intermediate states.

\begin{figure*}
  \centering
  \includegraphics[width=0.7\columnwidth]{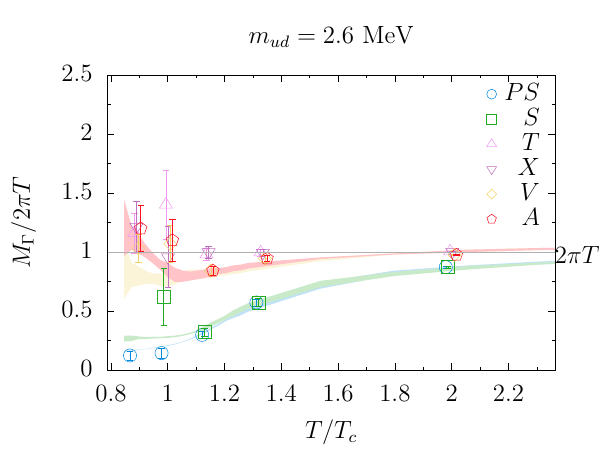}
  \caption{
The same figure as Fig.~\ref{fig:mscreenT} but the temperature is normalized by $T_c=165$ MeV.
and the screening mass is normalized by $2\pi T$.
For a comparison, the results by HotQCD collaboration \cite{Bazavov:2019www} which simulated  $N_f=2+1$ 
HISQ quarks are shown by the shadows. Except for the $S$ channels at the lowest two simulated temperatures,
the two results look consistent with each other.
  }
  \label{fig:mscreenTrescaled}
\end{figure*}

The same plot but with the temperature normalized by $T_c=165$ MeV 
and the screening mass normalized by $2\pi T$ is shown in Fig.~\ref{fig:mscreenTrescaled}.
For a comparison, the results by HotQCD collaboration \cite{Bazavov:2019www} which simulated  $N_f=2+1$ 
HISQ quarks are shown by the shadows(normalized by $2\pi T$ and the 2+1-flavor $T_c$ estimate respectively).
Although the value of $T_c$ is different due to the absence/presence of the strange quark,
the two results appear consistent and share many of the same features.
The only exception is the $S$ data at the lowest two simulated temperatures;
as explained in \cite{Bazavov:2019www}, the shadow approaches the twice of the pion mass
in the $T=0$ limit, due to a lattice artifact of the staggered quark action.
In our M\"obius domain-wall fermion formulation, this never happens
since the transition from the $S$ triplet to two $PS$ mesons 
is prohibited by the exact isospin symmetry.

\begin{figure*}[phtb]
  \centering
  \includegraphics[width=0.7\columnwidth]{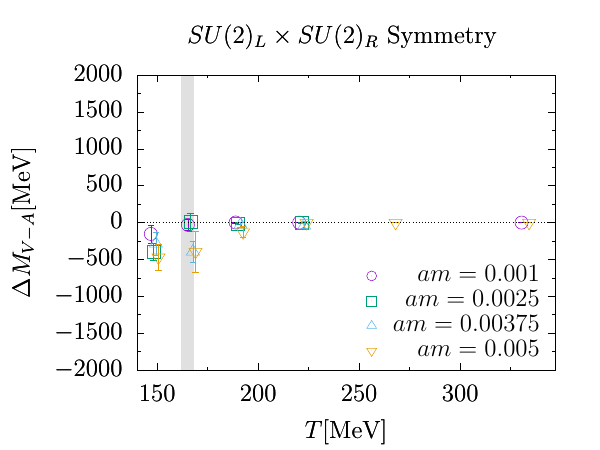}
  \caption{
    Temperature $T$ vs. the screening mass difference between $V$ and $A$ which probes the
    $SU(2)_L\times SU(2)_R$ symmetry. The vertical grey band indicates our estimate for the critical temperature 
$T_c=165(3)$ MeV from the 
chiral susceptibility.
  }
  \label{fig:symSU2}
  \centering
  \includegraphics[width=0.7\columnwidth]{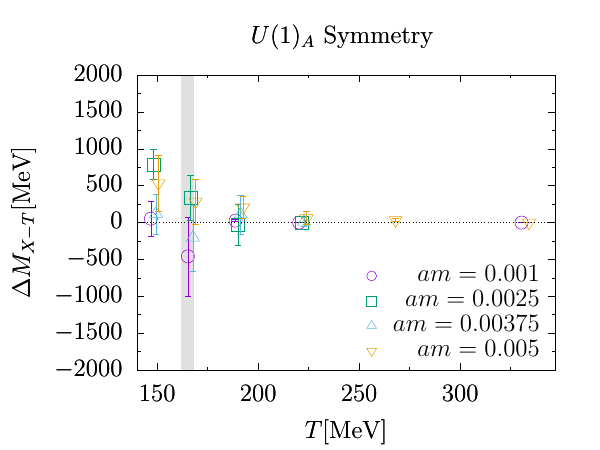}
  \caption{
    Temperature $T$ vs. the screening mass difference between $X_t$ and $T_t$ which probes the
    $U(1)_A$ symmetry. The vertical grey band indicates our estimate for the critical temperature $T_c=165(3)$ MeV from the 
chiral susceptibility.
  }
  \label{fig:symU1}
\end{figure*}

In Fig.~\ref{fig:symSU2}, we plot the mass difference between
$A$ and $V$ channels $\Delta m_{V-A}$ , which probes the $SU(2)_L \times SU(2)_R$ chiral
symmetry breaking, as a function of temperature $T$.
Here the data from the largest spatial volume for each simulated temperature is plotted.
Around the critical temperature shown by the vertical grey band,
the mass difference $\Delta m_{V-A}$ quickly vanishes.
This is most clearly seen at our lightest mass $m=0.001$, 
we find $\Delta m_{V-A}/m_A\sim 0.1$\%  at $T=190$ MeV
and 2.8\% at $T=165$ MeV. Further evidence of this rapid drop off is also seen in the literature with a rapid suppression of several order parameters associated with chiral symmetry shown in \cite{Krasniqi:2024kwm}.
For the higher quark masses, $\Delta m_{V-A}$ shows a milder 
$T$-dependence, which suggests that the crossover temperature becomes higher.
Both values are consistent with zero to within a standard deviation.

\begin{figure*}
  \centering
  \includegraphics[width=0.7\columnwidth]{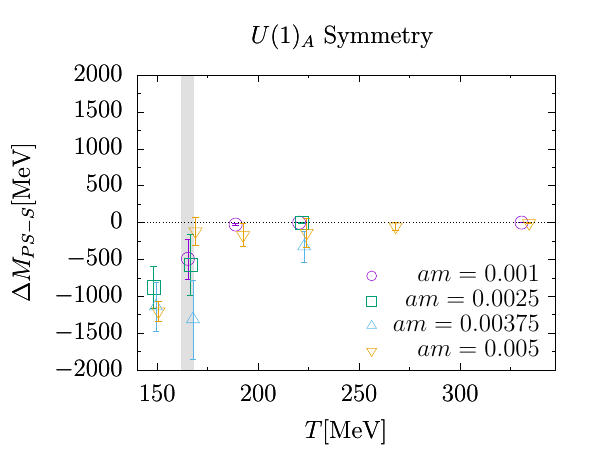}
  \caption{
    Temperature $T$ vs. the screening mass difference between $PS$ and $S$ which probes the
    $U(1)_A$ symmetry. The vertical grey band indicates our estimate for the critical temperature $T_c=165(3)$ MeV from the 
chiral susceptibility.
  }
  \label{fig:symU12}
  \centering
  \includegraphics[width=0.7\columnwidth]{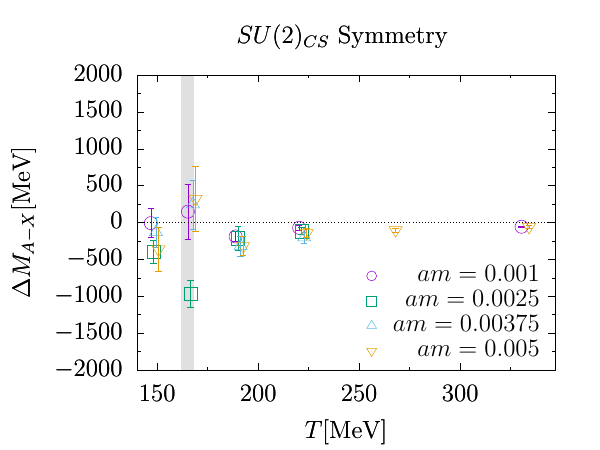}
  \caption{
    Temperature $T$ vs. the screening mass difference between $A$ and $X_t$ which probes the
    $SU(2)_{CS}$ symmetry. The vertical grey band indicates our estimate for the critical temperature $T_c=165(3)$ MeV from the 
chiral susceptibility.
  }
  \label{fig:symCS}
\end{figure*}

A similar behavior but with larger errorbars is found
in Figs.~\ref{fig:symU1} and \ref{fig:symU12} where the
mass differences between $X_t$ and $T_t$,  $\Delta m_{X-T}$, and
that of $PS$ and $S$, $\Delta m_{PS-S}$, are plotted.
These are the probes for the axial $U(1)_A$ symmetry.
We find that $\Delta m_{X-T}/m_X$ is 2.5\% at $T=190$ MeV
and 12.7\% at $T=165$ MeV\footnote{
In recent study \cite{Gavai:2024mcj} with M\"obius domain-wall quarks, 
violation of the axial $U(1)_A$ symmetry even at $T=186$ MeV was reported.
We notice, however, that their lattice spacings $a>0.13$ fm are coarser
than the one $a\sim 0.1$ fm we simulated in our previous work \cite{Cossu:2015kfa}
where we find a larger violation of the Ginsparg-Wilson in the low-lying
eigenmodes of the Dirac operator than the residual mass, 
which leads to overestimation of the axial $U(1)_A$ anomaly.
}.
The latter is consistent with zero within a larger errorbar.
This indicates that the chiral anomaly effect is
quite suppressed above $T_c$.

\begin{figure*}
  \centering
  \includegraphics[width=0.7\columnwidth]{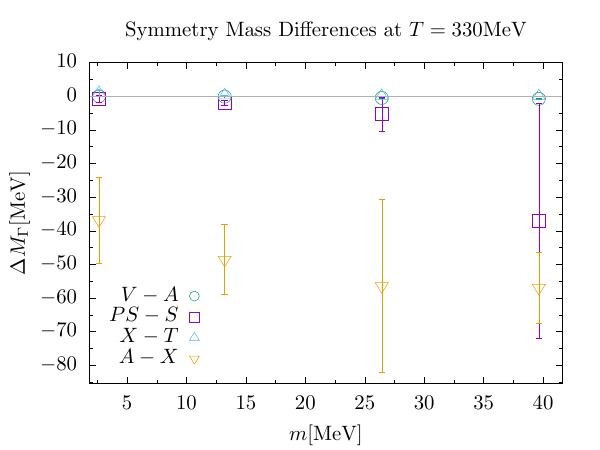}
  \caption{
 Quark mass dependence of $\Delta m_{V-A}$ (circles), $\Delta m_{PS-S}$ (squares), 
$\Delta m_{X-T}$ (upward triangles), and $\Delta m_{A-X}$(downward triangles) at $T=330$ MeV.
Compared to the former three probing $SU(2)_L\times SU(2)_R$ and  $U(1)_A$ symmetries,
which are consistent with zero with errorrbars $\sim 1$ MeV, 
the chiral-spin probe $\Delta m_{A-X}$ is clearly nonzero: $\sim -40$ MeV
and the associated chiral-spin symmetry $SU(2)_{CS}$ is only approximate:
$|\Delta m_{X-A}|/T \sim 0.12$.
  }
  \label{fig:symCS330}
\end{figure*}

Fig.~\ref{fig:symCS} is a plot of the chiral-spin $SU(2)_{CS}$ symmetry, 
which is an approximate symmetry expected up to $O(1/T)$ corrections, as
explained in Sec.~\ref{sec:corr}. This symmetry 
is examined by the mass difference $\Delta m_{X-A}$ between $X_t$ and $A$.
The $T$ dependence for the screening mass, in addition to the reduction in noise, is analogous to $SU(2)_L\times SU(2)_R$ and $U(1)_A$;
however, we find a qualitative difference of $SU(2)_{CS}$ in Fig.~\ref{fig:symCS330}. Here we plot the quark mass dependence of 
$\Delta m_{V-A}$ (circles), $\Delta m_{PS-S}$ (squares), 
$\Delta m_{X-T}$ (upward triangles), and $\Delta m_{A-X}$ (downward triangles) at $T=330$ MeV.
The former three probing $SU(2)_L\times SU(2)_R$ and  $U(1)_A$ symmetries,
are consistent with zero, and their uncertainty is $\sim 1$MeV, 
which is less than 1\% of $T$.
In contrast to these other mass differences, the data $\Delta m_{X-A}$ is clearly nonzero even at the lightest mass: $\sim -40$ MeV
showing that the associated chiral-spin symmetry $SU(2)_{CS}$ is only approximate:
$|\Delta m_{X-A}|/T \sim 0.12$.
We note that at much higher temperatures than $T=1$ GeV, 
it was reported in Ref.~\cite{DallaBrida:2021ddx} that the mass difference between the $PS$ and $V$ channels goes down to $\sim 1$\% of $2\pi T$ although, the difference is still statistically significant and deviates from the next-to-leading order perturbative QCD prediction,
where the one-loop correction of each meson mass to $2\pi T$ is positive and channel-independent
(for related results and discussion see \cite{Laudicina:2022ghk, Giusti:2024ohu}). 

\begin{figure*}[phtb]
  \centering
  \includegraphics[width=0.49\columnwidth]{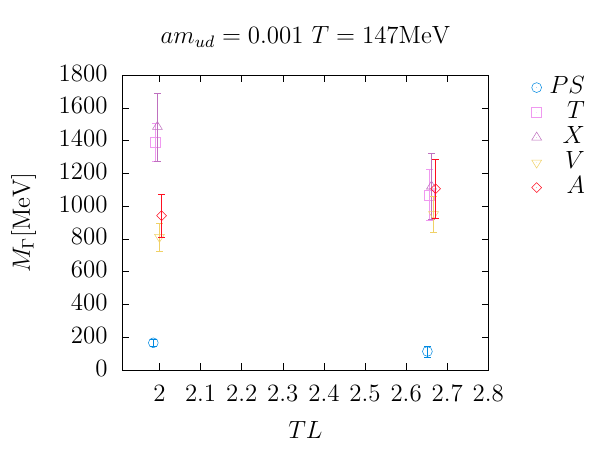}
  \includegraphics[width=0.49\columnwidth]{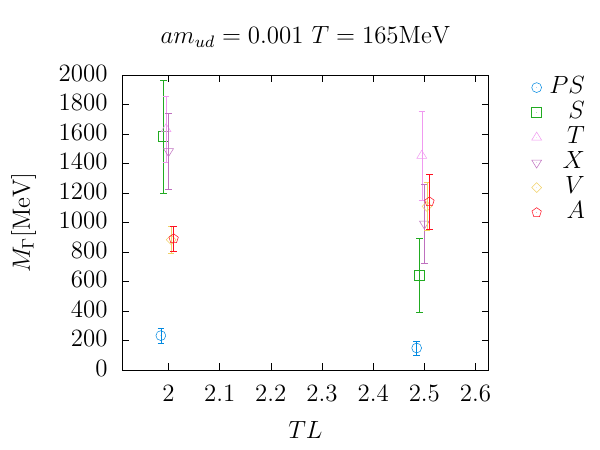}
  \includegraphics[width=0.49\columnwidth]{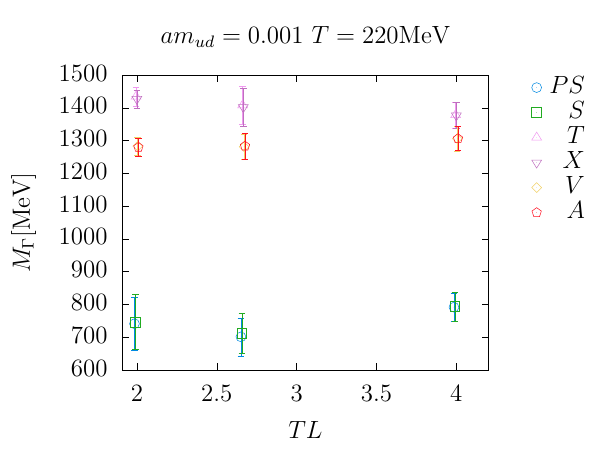}
  \caption{
 The lattice size $L$ vs. screening masses at $T=147$, 165 and 220 MeV 
 is plotted. We do not see any significant finite volume dependence beyond $2\sigma$.
  }
  \label{fig:finiteV}
\end{figure*}

Finally let us discuss possible systematics in our numerical results.
In Fig.~\ref{fig:finiteV}, we plot the lattice size $L$ dependence of all measured 
screening masses at $T=147$, 165 and 220 MeV. 
The data at different $L$ are consistent within two standard deviations
and we do not see any significant finite volume dependence.
Due to our fixed lattice spacing at $a=0.075$ fm, we 
cannot numerically estimate the discretization effects.
Our choice is relatively fine compared to other groups
and we expect the automatic $O(a)$ improvement due to 
the chiral symmetry of the M\"obius domain-wall fermion action.
Note also, that our simulated up and down quark mass range covers the physical point,
and the lightest quark mass is sufficiently close to the chiral limit.


\section{Summary and discussion}
\label{sec:summary}

We have simulated $N_f=2$ lattice QCD at finite temperatures in the range $[0.9,2]T_c$
with the M\"obius domain-wall quark action and Symanzik gauge action
setting the lattice spacing at $a\sim 0.075$ fm. 
Our simulated quark mass covers the physical point, 
and thus the chiral symmetry is well under control with 
the residual mass $\sim 0.14$ MeV.
For the lowest two temperatures, we have simulated multiple volumes
in order to check the finite size systematics.

The two-point correlation functions of the isovector operators 
with different spins have been measured in the spatial directions.
From the standard $\cosh$ fit, we have extracted the screening masses for the six channels of the isospin triplet.
The temperature dependence of the screening masses shows 
several remarkable features.
First, the values at the lightest simulated quark mass $m=0.001$ at $T=0.9T_c$
are consistent with the experimental values at $T=0$.
This may indicate that the chiral symmetry is strongly broken, even for high temperatures below the psuedocritical point.
Second, the scalar $S$, axial vector $A$ and axial tensor $X_t$
masses all reduce above $T\sim T_c$ to values similar to
their $SU(2)_L \times SU(2)_R$ or $U(1)_A$ partners $PS$, $V$ and $T_t$.
Then above $T_c$, all the channels monotonically increase
and appear to converge to twice the ground state Matsubara mass $2\pi T$.

Taking the difference of the screening masses between the aforementioned channel partners,
we have examined various symmetries of QCD at high temperatures.
The standard chiral $SU(2)_L \times SU(2)_R$ is restored at the critical temperature $T_c\sim 165$MeV (observable from the degeneracy of the $V$ and $A$ channels), consistent with the critical temperature estimated from the chiral susceptibility in previous work.
The axial $U(1)_A$ symmetry behaves similarly, with degeneracy between the $T_t$ and $X_t$ as well as $PS$ and $S$ at or close to $T_c$ although the signals are noisier.
The mass difference between these three channel pairs at $T\sim 2T_c$ is consistent with zero
with quite small uncertainty less than 1 MeV, or 1\% of the temperature.
In contrast to the above two symmetries, 
the chiral spin $SU(2)_{CS}$ probed by $A$ and $X_t$ is only approximate at $T\sim 2T_c$,
showing a breaking $\sim 40$ MeV, which is 12\% of the temperature.


\begin{acknowledgments}
We thank T.~Cohen, H.-T.~Ding, C.~Gattringer, A.~Hasenfratz, C.B.~Lang,
O.~Phillipsen, R.~Pisarski, S.~Prelovsek, for fruitful discussions.
Discussions during the long-term workshop, HHIQCD2024, at Yukawa Institute for
Theoretical Physics (YITP-T-24-02) were useful to complete this work.
We also thank the members of JLQCD collaboration for their encouragement
and support.
 Numerical simulations are performed on IBM System Blue Gene Solution at KEK under a support of its Large Scale Simulation Program (No.
16/17-14), Oakforest-PACS at JCAHPC under a support of the HPCI System Research Projects
(Project IDs: hp170061, hp180061, hp190090, hp200086, and hp210104) and Multidisciplinary
Cooperative Research Program in CCS, University of Tsukuba (xg17i032 and xg18i023), the
supercomputer Fugaku provided by the RIKEN Center for Computational Science (hp200130,
hp210165, hp210231, hp220279 and hp230323), Wisteria/BDEC-01 Odyssey at JCAHPC (HPCI:
hp220093, hp230070 and hp240050	 MCRP: wo22i038), and Polarie and Grand Chariot at Hokkaido University (hp200130). 
We used Japan Lattice Data Grid (JLDG) for storing a part of the numerical
data generated for this work.
This work is supported in part by the Japanese Grant-in-Aid for Scientific
Research (No. JP26247043, JP18H01216 and JP18H04484, JP22H01219), and by MEXT as ``Priority Issue on
Post-K computer'' (Elucidation of the Fundamental Laws and Evolution of the Universe) and by
Joint Institute for Computational Fundamental Science (JICFuS).

\end{acknowledgments}

\begin{table}[tbh]
\centering
\scriptsize
\renewcommand{\arraystretch}{0.95}
\begin{tabular}{c c c c c c c | c c c c c c }
   \hline\hline
   $L^3\times L_t$ & $L$(fm) & $T\mbox{[MeV]}$ & $TL$ &  am      &  $m\mbox{[MeV]}$& No. samples &    $m_{S}$ &   $m_{PS}$&   $m_{V}$ &   $m_{A}$ &   $m_{X}$ &   $m_{T}$\\
   \hline
   $48^3\times 18$ & 3.6     & 147             & 2.7  &  0.00100 &  2.6            & 146 &            &    113(34)&   951(111) &  1106(182) &  1120(201) &  1068(154)\\
                   &         &                 &      &  0.00250 &  6.6            & 40  &  1097(288) &     220(5)&   1009(69) &  1404(129) &  1801(166) &   1012(71)\\
                   &         &                 &      &  0.00375 &  9.9            & 40  &  1387(337) &     245(5)&    854(74) &  1145(155) &  1279(223) &   1164(97)\\
                   &         &                 &      &  0.00500 &  13.2           & 83  &            &     135(7)&    903(58) &  1375(177) &  1733(281) &  1199(117)\\
   $36^3\times 18$ & 2.7     & 147             & 2.0  &  0.00100 &  2.6            & 146 &            &    166(22)&    812(85) &   941(132) &  1483(206) &  1391(115)\\
                   &         &                 &      &  0.00250 &  6.6            & 121 &  2605(533) &    220(12)&    767(53) &   930(156) &  1645(153) &  1228(117)\\
                   &         &                 &      &  0.00375 &  9.9            & 122 &  1586(333) &    240(11)&    886(77) &  1046(181) &  1635(107) &   1216(52)\\
                   &         &                 &      &  0.00500 &  13.2           & 131 &  1490(139) &     288(6)&    861(43) &  1231(125) &  1527(117) &   1091(56)\\
   \hline
   $40^3\times 16$ & 3.0     & 165             & 2.5  &  0.00100 &  2.6            & 165 &   642(251) &    149(46)&  1109(165) &  1141(185) &   994(268) &  1453(305)\\
                   &         &                 &      &  0.00250 &  6.6            & 95  &   817(402) &    245(18)&   544(146) &   536(201) &  1499(196) &  1161(146)\\
                   &         &                 &      &  0.00375 &  9.9            & 97  &            &    301(16)&   1003(90) &  1400(158) &  1156(305) &  1364(171)\\
                   &         &                 &      &  0.00500 &  13.2           & 95  &            &    273(20)&   936(113) &  1330(216) &  1007(341) &    721(95)\\
   $32^3\times 16$ & 2.4     & 165             & 2.0  &  0.00100 &  2.6            & 165 &            &    233(49)&    883(93) &    890(84) &  1485(256) &  1633(222)\\
                   &         &                 &      &  0.00250 &  6.6            & 116 &            &    362(53)&    904(75) &   978(133) &  1268(431) &  1212(176)\\
                   &         &                 &      &  0.00375 &  9.9            & 163 &  1624(538) &    309(14)&    934(90) &  1047(131) &  1552(178) &   966(152)\\
                   &         &                 &      &  0.00500 &  13.2           & 143 &   444(187) &    326(24)&    932(84) &  1106(159) &  1679(207) &   1216(89)\\
   \hline
   $32^3\times 14$ & 2.4     & 189             & 2.3  &  0.00100 &  2.6            & 190 &    381(40) &    353(39)&    998(40) &    998(41) &   1187(62) &   1161(61)\\
                   &         &                 &      &  0.00250 &  6.6            & 177 &            &    362(39)&   1150(49) &   1164(52) &  1376(154) &  1407(131)\\
                   &         &                 &      &  0.00375 &  9.9            & 137 &            &    402(27)&    964(33) &    996(44) &  1318(146) &  1212(135)\\
                   &         &                 &      &  0.00500 &  13.2           & 133 &   582(159) &    413(29)&    940(44) &   1069(74) &  1385(114) &   1178(82)\\
   \hline
   $48^3\times 12$ & 3.6     & 220             & 4.0  &  0.00100 &  2.6            & 220 &    793(43) &    792(43)&   1306(37) &   1306(37) &   1377(39) &   1378(39)\\
                   &         &                 &      &  0.00250 &  6.6            & 97  &    913(21) &    911(20)&   1290(39) &   1293(39) &   1411(38) &   1410(38)\\
                   &         &                 &      &  0.00375 &  9.9            & 114 &  1033(192) &    706(47)&   1189(78) &   1235(93) &   1441(48) &   1431(54)\\
                   &         &                 &      &  0.00500 &  13.2           & 116 &   888(108) &   752(109)&   1234(33) &   1236(35) &   1378(58) &   1316(48)\\
   $40^3\times 12$ & 3.0     & 220             & 3.3  &  0.00500 &  13.2           & 220 &  1534(351) &    502(51)&   1327(32) &   1298(28) &  1291(116) &   1425(38)\\
                   &         &                 &      &  0.01000 &  26.4           & 244 &  1495(336) &    696(35)&   1222(24) &   1275(32) &  1302(106) &   1444(99)\\
   $32^3\times 12$ & 2.4     & 220             & 2.7  &  0.00100 &  2.6            & 532 &    712(61) &    701(58)&   1281(38) &   1283(39) &   1402(58) &   1408(58)\\
                   &         &                 &      &  0.00250 &  6.6            & 534 &    697(88) &    797(85)&   1235(22) &   1234(23) &   1294(46) &   1307(51)\\
                   &         &                 &      &  0.00375 &  9.9            & 689 &   615(158) &    696(77)&   1244(23) &   1251(24) &  1570(103) &  1374(104)\\
                   &         &                 &      &  0.00500 &  13.2           & 544 &  1350(360) &    717(90)&   1182(25) &   1213(23) &  1465(115) &   1292(69)\\
                   &         &                 &      &  0.01000 &  26.4           & 622 &  1254(226) &    686(36)&   1227(57) &   1160(76) &  1345(101) &   1302(80)\\
   $24^3\times 12$ & 1.8     & 220             & 2.0  &  0.00100 &  2.6            & 373 &    747(84) &    741(82)&   1283(27) &   1280(29) &   1427(28) &   1434(28)\\
                   &         &                 &      &  0.00250 &  6.6            & 361 &  1637(334) &   806(121)&   1292(30) &   1295(30) &   1345(77) &   1305(80)\\
                   &         &                 &      &  0.00375 &  9.9            & 331 &   687(120) &    785(53)&   1333(32) &   1335(33) &   1392(32) &   1398(32)\\
                   &         &                 &      &  0.00500 &  13.2           & 363 &  2505(488) &   736(134)&   1219(31) &   1245(36) &   1327(66) &   1469(55)\\
                   &         &                 &      &  0.01000 &  26.4           & 365 &   770(153) &    713(54)&   1271(29) &   1307(38) &   1421(33) &   1450(29)\\
   \hline
   $32^3\times 10$ & 2.4     & 264             & 3.2  &  0.00500 &  13.2           & 640 &   1319(35) &   1266(26)&   1566(18) &   1566(18) &   1666(32) &   1631(26)\\
                   &         &                 &      &  0.00800 &  21.1           & 237 &   1252(19) &   1251(19)&   1572(15) &   1574(15) &   1626(19) &   1627(20)\\
                   &         &                 &      &  0.01000 &  26.4           & 291 &  1525(121) &   1117(94)&   1560(19) &   1581(27) &   1630(49) &   1654(28)\\
                   &         &                 &      &  0.01500 &  39.6           & 121 &  1458(123) &  1236(100)&   1596(21) &   1599(21) &   1666(28) &   1693(29)\\
   \hline
   $32^3\times 8$  & 2.4     & 330             & 4.0  &  0.00100 &  2.6            & 260 &   1810(11) &   1809(11)&    2026(9) &    2026(9) &   2083(13) &   2083(13)\\
                   &         &                 &      &  0.00500 &  13.2           & 317 &   1796(19) &   1791(18)&   2038(12) &   2038(12) &   2094(24) &   2094(24)\\
                   &         &                 &      &  0.01000 &  26.4           & 350 &   1783(15) &   1781(15)&    2033(8) &    2033(8) &    2082(9) &    2082(9)\\
                   &         &                 &      &  0.01500 &  39.6           & 306 &   1828(24) &   1791(22)&   2027(10) &   2028(10) &   2064(17) &   2063(17)\\
                   &         &                 &      &  0.02000 &  52.9           & 218 &   1807(15) &   1796(14)&   2013(18) &   2014(18) &   2069(15) &   2066(15)\\
                   &         &                 &      &  0.04000 &  105.7          & 164 &   1824(27) &   1770(20)&   2021(16) &   2024(16) &   2080(16) &   2073(15)\\
   \hline
\end{tabular}

\caption{Simulation parameters and meson screening mass in the physical unit [MeV] obtained from the
$\cosh$ fit.\label{tab:parameters-results}}
\end{table}

\begin{table}[tbh]
\centering
\renewcommand{\arraystretch}{0.95}
\scriptsize
\begin{tabular}{c c c c c c c | c c c c }
   \hline\hline
   $L^3\times L_t$ & $L$(fm) &  $T\mbox{[MeV]}$ &  $TL$ & am      &  $m\mbox{[MeV]}$ & No. samples &  $m_{PS-S}$ &  $m_{V-A}$  &  $m_{X-T}$  & $m_{A-X}$\\
   \hline                                                                                   
   $48^3\times 18$ & 3.6     &  147             &  2.7  & 0.00100 &  2.6             & 146 &             &  -155(120)  &    52(233)  &    -8(196)\\
                   &         &                  &       & 0.00250 &  6.6             & 40  &  -877(289)  &  -395(119)  &   789(201)  &  -398(158)\\
                   &         &                  &       & 0.00375 &  9.9             & 40  & -1141(335)  &  -291(155)  &   115(274)  &  -133(201)\\
                   &         &                  &       & 0.00500 &  13.2            & 83  &             &  -472(179)  &   534(381)  &  -361(297)\\
   $36^3\times 18$ & 2.7     &  147             &  2.0  & 0.00100 &  2.6             & 146 &             &   -129(66)  &    92(290)  &  -539(213)\\
                   &         &                  &       & 0.00250 &  6.6             & 121 & -2385(525)  &  -163(139)  &   417(235)  &  -710(169)\\
                   &         &                  &       & 0.00375 &  9.9             & 122 & -1346(331)  &  -160(171)  &   419(127)  &  -586(114)\\
                   &         &                  &       & 0.00500 &  13.2            & 131 & -1203(141)  &  -370(127)  &   437(147)  &  -296(144)\\
   \hline                                                                                   
   $40^3\times 16$ & 3.0     &  165             &  2.5  & 0.00100 &  2.6             & 165 &  -493(272)  &    -32(75)  &  -459(536)  &   148(377)\\
                   &         &                  &       & 0.00250 &  6.6             & 95  &  -572(414)  &     8(121)  &   338(307)  &  -963(184)\\
                   &         &                  &       & 0.00375 &  9.9             & 97  &             &  -397(143)  &  -208(447)  &   244(329)\\
                   &         &                  &       & 0.00500 &  13.2            & 95  &             &  -394(279)  &   287(304)  &   323(439)\\
   $32^3\times 16$ & 2.4     &  165             &  2.0  & 0.00100 &  2.6             & 165 &             &     -7(28)  &  -149(464)  &  -595(281)\\
                   &         &                  &       & 0.00250 &  6.6             & 116 &             &   -74(151)  &    56(542)  &  -290(445)\\
                   &         &                  &       & 0.00375 &  9.9             & 163 & -1315(539)  &  -113(128)  &   586(247)  &  -505(248)\\
                   &         &                  &       & 0.00500 &  13.2            & 143 &  -118(194)  &  -174(138)  &   463(221)  &  -573(227)\\
   \hline                                                                                   
   $32^3\times 14$ & 2.4     &  189             &  2.3  & 0.00100 &  2.6             & 190 &    -28(14)  &       0(4)  &     26(16)  &   -188(67)\\
                   &         &                  &       & 0.00250 &  6.6             & 177 &             &    -14(16)  &   -30(274)  &  -213(159)\\
                   &         &                  &       & 0.00375 &  9.9             & 137 &             &    -32(35)  &   107(265)  &  -321(140)\\
                   &         &                  &       & 0.00500 &  13.2            & 133 &  -169(159)  &   -129(75)  &   207(144)  &  -315(127)\\
   \hline                                                                                   
   $48^3\times 12$ & 3.6     &  220             &  4.0  & 0.00100 &  2.6             & 220 &      -2(1)  &      -0(0)  &      -1(0)  &    -71(37)\\
                   &         &                  &       & 0.00250 &  6.6             & 97  &      -2(1)  &      -3(1)  &       0(1)  &   -118(47)\\
                   &         &                  &       & 0.00375 &  9.9             & 114 &  -327(206)  &    -46(35)  &     10(57)  &   -206(79)\\
                   &         &                  &       & 0.00500 &  13.2            & 116 &  -136(196)  &     -2(13)  &     62(86)  &   -141(62)\\
   $40^3\times 12$ & 3.0     &  220             &  3.3  & 0.00500 &  13.2            & 220 & -1032(399)  &     14(18)  &    -22(96)  &   -141(98)\\
                   &         &                  &       & 0.01000 &  26.4            & 244 &  -799(347)  &    -48(20)  &     -2(79)  &    -87(79)\\
   $32^3\times 12$ & 2.4     &  220             &  2.7  & 0.00100 &  2.6             & 532 &     -11(7)  &      -2(2)  &      -5(4)  &   -119(80)\\
                   &         &                  &       & 0.00250 &  6.6             & 534 &   100(102)  &       1(2)  &    -13(32)  &    -60(36)\\
                   &         &                  &       & 0.00375 &  9.9             & 689 &    81(109)  &      -6(2)  &   197(195)  &   -320(98)\\
                   &         &                  &       & 0.00500 &  13.2            & 544 &  -633(444)  &    -31(17)  &   173(135)  &  -252(112)\\
                   &         &                  &       & 0.01000 &  26.4            & 622 &  -568(243)  &     67(79)  &    44(131)  &  -185(124)\\
   $24^3\times 12$ & 1.8     &  220             &  2.0  & 0.00100 &  2.6             & 373 &      -6(4)  &       4(4)  &      -7(5)  &   -148(38)\\
                   &         &                  &       & 0.00250 &  6.6             & 361 &  -831(225)  &      -3(2)  &     40(67)  &    -48(97)\\
                   &         &                  &       & 0.00375 &  9.9             & 331 &     98(74)  &      -1(1)  &      -6(6)  &    -56(43)\\
                   &         &                  &       & 0.00500 &  13.2            & 363 & -1769(375)  &    -26(15)  &  -144(111)  &    -81(78)\\
                   &         &                  &       & 0.01000 &  26.4            & 365 &   -57(114)  &    -36(23)  &    -28(36)  &   -115(45)\\
   \hline                                                                                   
   $32^3\times 10$ & 2.4     &  264             &  3.2  & 0.00500 &  13.2            & 640 &    -53(53)  &      -1(1)  &     35(18)  &   -100(28)\\
                   &         &                  &       & 0.00800 &  21.1            & 237 &      -0(3)  &      -1(1)  &      -0(3)  &    -53(20)\\
                   &         &                  &       & 0.01000 &  26.4            & 291 &  -407(203)  &    -21(26)  &    -25(67)  &    -49(69)\\
                   &         &                  &       & 0.01500 &  39.6            & 121 &  -222(217)  &      -3(1)  &    -27(28)  &    -67(30)\\
   \hline                                                                                   
   $32^3\times 8$  & 2.4     &  330             &  4.0  & 0.00100 &  2.6             & 260 &      -1(1)  &      -0(0)  &      -0(0)  &    -57(11)\\
                   &         &                  &       & 0.00500 &  13.2            & 317 &      -5(5)  &      -0(0)  &       0(0)  &    -56(26)\\
                   &         &                  &       & 0.01000 &  26.4            & 350 &      -2(1)  &      -0(0)  &       0(0)  &    -49(10)\\
                   &         &                  &       & 0.01500 &  39.6            & 306 &    -37(35)  &      -1(0)  &       1(1)  &    -37(13)\\
                   &         &                  &       & 0.02000 &  52.9            & 218 &     -12(6)  &      -2(0)  &       2(1)  &    -54(12)\\
                   &         &                  &       & 0.04000 &  105.7           & 164 &    -54(30)  &      -4(2)  &       6(3)  &    -55(11)\\
   \hline
\end{tabular}

\caption{The difference between the screening masses of the mesons in the physical unit [MeV]. \label{tab:sym-results}}
\end{table}

\begin{appendix}
\section{Raw correlators}
\begin{figure*}[t]
\centering
\includegraphics[width=0.45\columnwidth]{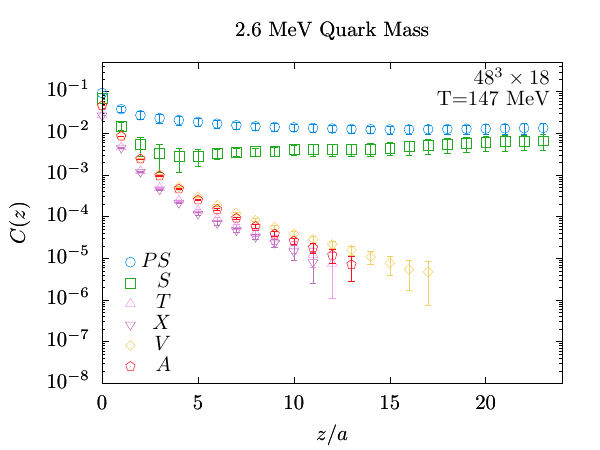}
\includegraphics[width=0.45\columnwidth]{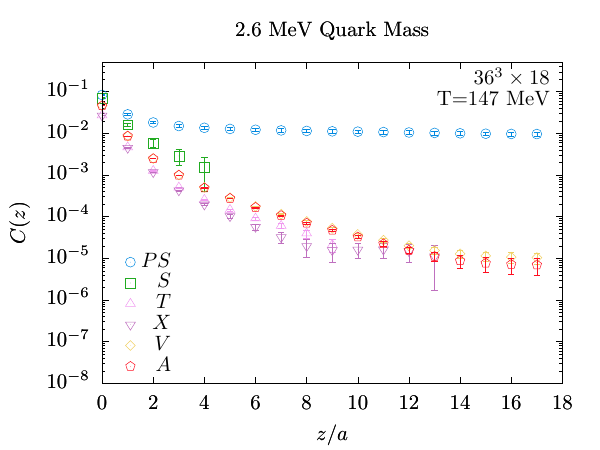}
\includegraphics[width=0.45\columnwidth]{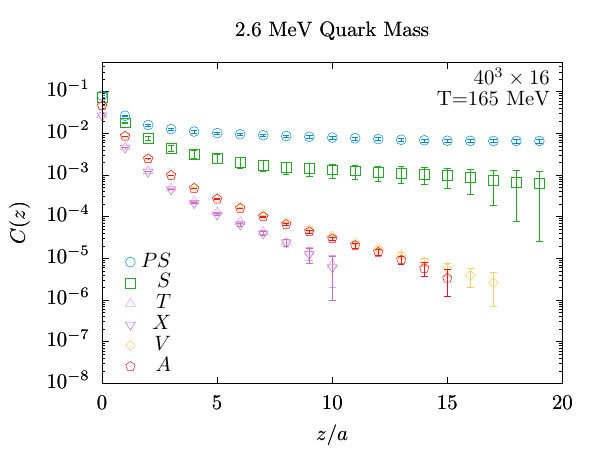}
\includegraphics[width=0.45\columnwidth]{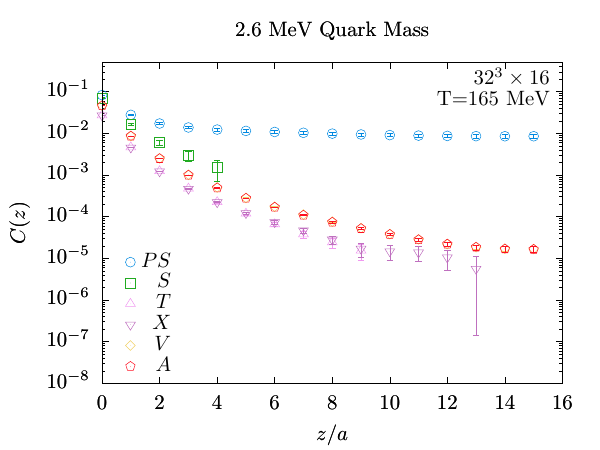}
\caption{Plots of the raw correlators for $T=147$ MeV and $165$ MeV. All plot of the correlators correlators are averaged over the three spatial directions and symmetrized with respect to the reflection of the axis $z\to -z$. \label{fig:147-165-corrplot}}
\end{figure*}

\begin{figure*}[b!]
\centering
\includegraphics[width=0.45\columnwidth]{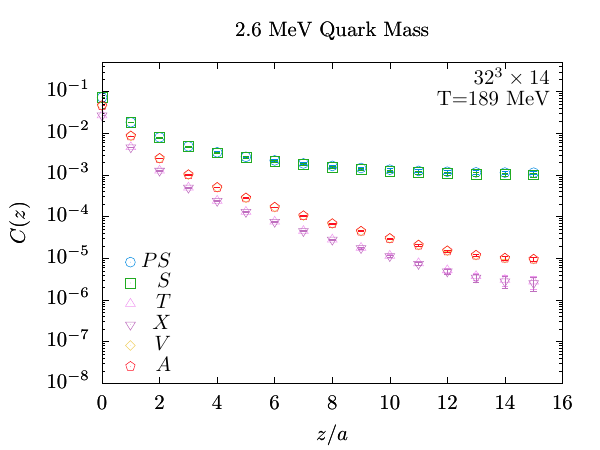}
\includegraphics[width=0.45\columnwidth]{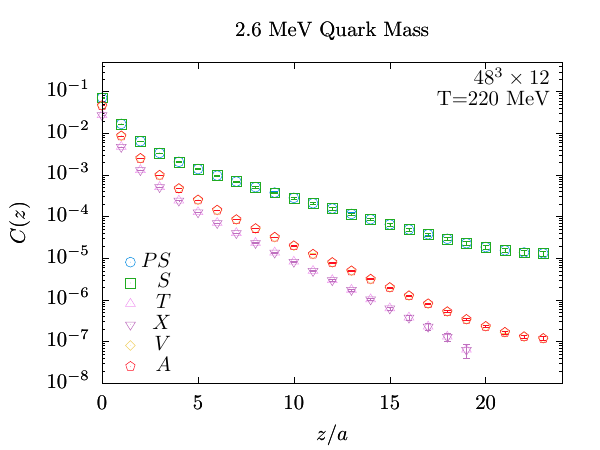}
\includegraphics[width=0.45\columnwidth]{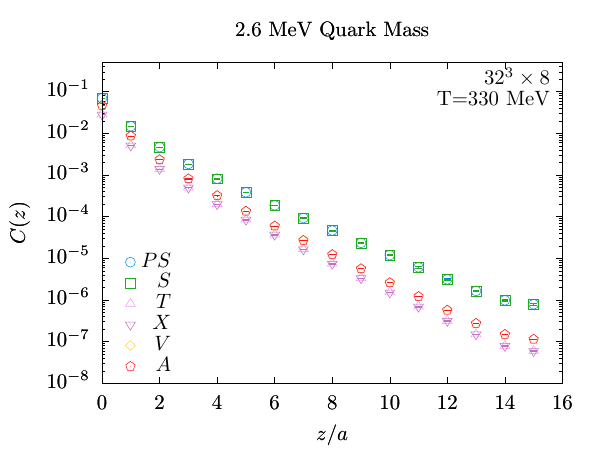}
\caption{The three higher temperature raw correlator plots like Fig.\ref{fig:147-165-corrplot}. These plots are also symmetrized around the midpoint of the correlator. \label{fig:189-330-corrplot}}
\end{figure*}

In this appendix, we summarize the results for the raw correlator
with the lightest quark mass $m=0.001(2.6 \mbox{MeV})$ 
at the simulated temperatures.
Some data having negative values are neglected. Correlators for two volumes at the lowest temperatures in our study are shown in figure \ref{fig:147-165-corrplot}, while the largest aspect ratio plots for temperatures above $T_c$ are shown in figure \ref{fig:189-330-corrplot}.

\section{Screening Mass Plots for $am>0.0010$}
\begin{figure*}[h]
\centering
\includegraphics[width=0.45\columnwidth]{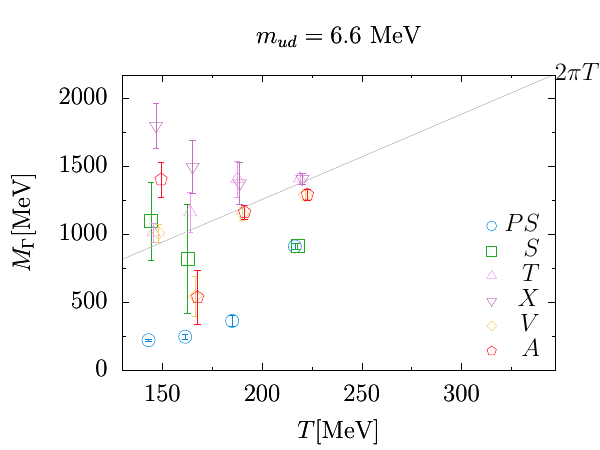}
\includegraphics[width=0.45\columnwidth]{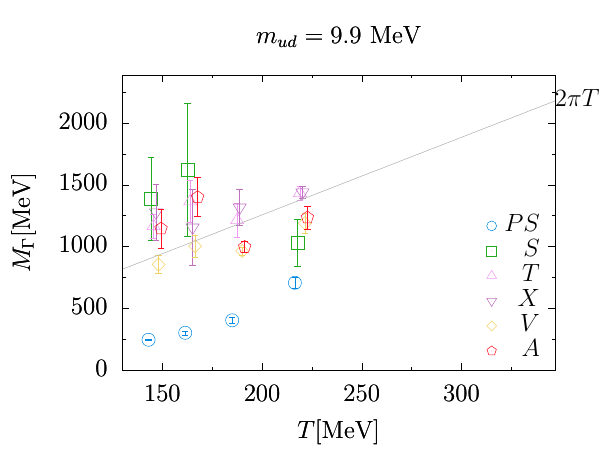}
\includegraphics[width=0.45\columnwidth]{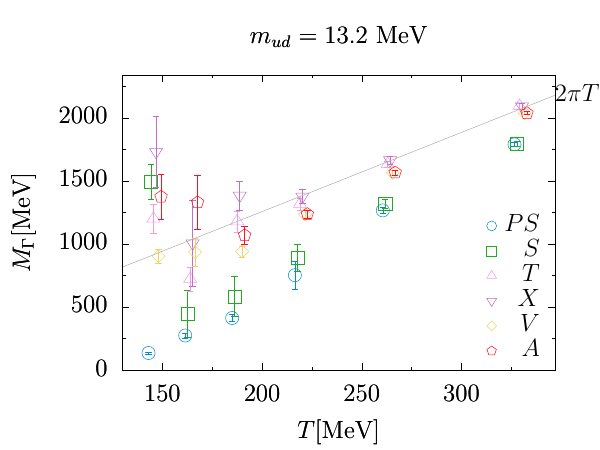}
\caption{These plots correspond to the three other quark mass ensembles above the physical point in our study $am=0.0025(6.6\mbox{MeV})$,$0.00375(9.9\mbox{MeV})$ and $0.0050(13.2\mbox{MeV}$). \label{fig:mscreenTmass}}
\end{figure*}

As an additional check of the quark mass dependence on the meson screening mass,
we present the same plots as Fig.~\ref{fig:mscreenT}  but with three different quark masses in Fig.~\ref{fig:mscreenTmass}.

\end{appendix}
\pagebreak
\raggedright


\begin{thebibliography}{99}


\bibitem{Pisarski:1983ms} 
  R.~D.~Pisarski and F.~Wilczek,
  ``Remarks on the Chiral Phase Transition in Chromodynamics,''
  Phys.\ Rev.\ D {\bf 29}, 338 (1984).
  doi:10.1103/PhysRevD.29.338
  

\bibitem{Gross:1980br}
D.~J.~Gross, R.~D.~Pisarski and L.~G.~Yaffe,
``QCD and Instantons at Finite Temperature,''
Rev. Mod. Phys. \textbf{53}, 43 (1981)
doi:10.1103/RevModPhys.53.43

\bibitem{Diakonov:1984vw}
D.~Diakonov and V.~Y.~Petrov,
``CHIRAL CONDENSATE IN THE INSTANTON VACUUM,''
Phys. Lett. B \textbf{147}, 351-356 (1984)
doi:10.1016/0370-2693(84)90132-1

\bibitem{Morris:1984zi}
T.~R.~Morris, D.~A.~Ross and C.~T.~Sachrajda,
``Higher Order Quantum Corrections in the Presence of an Instanton Background Field,''
Nucl. Phys. B \textbf{255}, 115-148 (1985)
doi:10.1016/0550-3213(85)90131-2

\bibitem{Schafer:1996wv}
T.~Sch\"afer and E.~V.~Shuryak,
``Instantons in QCD,''
Rev. Mod. Phys. \textbf{70}, 323-426 (1998)
doi:10.1103/RevModPhys.70.323
[arXiv:hep-ph/9610451 [hep-ph]].

\bibitem{Ringwald:1999ze}
A.~Ringwald and F.~Schrempp,
``Confronting instanton perturbation theory with QCD lattice results,''
Phys. Lett. B \textbf{459}, 249-258 (1999)
doi:10.1016/S0370-2693(99)00682-6
[arXiv:hep-lat/9903039 [hep-lat]].


\bibitem{Cohen:1996ng} 
  T.~D.~Cohen,
  ``The High temperature phase of QCD and U(1)-A symmetry,''
  Phys.\ Rev.\ D {\bf 54}, R1867 (1996)
  doi:10.1103/PhysRevD.54.R1867
  [hep-ph/9601216].

\bibitem{Cohen:1997hz} 
  T.~D.~Cohen,
  ``The Spectral density of the Dirac operator above T(c) rep,''
  nucl-th/9801061.

\bibitem{Aoki:2012yj} 
  S.~Aoki, H.~Fukaya and Y.~Taniguchi,
  ``Chiral symmetry restoration, eigenvalue density of Dirac operator and axial U(1) anomaly at finite temperature,''
  Phys.\ Rev.\ D {\bf 86}, 114512 (2012)
  doi:10.1103/PhysRevD.86.114512
  [arXiv:1209.2061 [hep-lat]].


\bibitem{Pelissetto:2013hqa}  
  A.~Pelissetto and E.~Vicari,
  ``Relevance of the axial anomaly at the finite-temperature chiral transition in QCD,''
  Phys.\ Rev.\ D {\bf 88}, no. 10, 105018 (2013)
  doi:10.1103/PhysRevD.88.105018
  [arXiv:1309.5446 [hep-lat]].
  
\bibitem{Kanazawa:2014cua} 
  T.~Kanazawa and N.~Yamamoto,
  ``Quasi-instantons in QCD with chiral symmetry restoration,''
  Phys.\ Rev.\ D {\bf 91}, 105015 (2015)
  doi:10.1103/PhysRevD.91.105015
  [arXiv:1410.3614 [hep-th]].

\bibitem{Sato:2014axa} 
  T.~Sato and N.~Yamada,
  ``Linking $U(2)\times U(2)$ to $O(4)$ model via decoupling,''
  Phys.\ Rev.\ D {\bf 91}, no. 3, 034025 (2015)
  doi:10.1103/PhysRevD.91.034025
  [arXiv:1412.8026 [hep-lat]].
  
\bibitem{Nakayama:2016jhq} 
  Y.~Nakayama and T.~Ohtsuki,
  ``Conformal Bootstrap Dashing Hopes of Emergent Symmetry,''
  Phys.\ Rev.\ Lett.\  {\bf 117}, no. 13, 131601 (2016)
  doi:10.1103/PhysRevLett.117.131601
  [arXiv:1602.07295 [cond-mat.str-el]].


\bibitem{Cossu:2013uua}
G.~Cossu, S.~Aoki, H.~Fukaya, S.~Hashimoto, T.~Kaneko, H.~Matsufuru and J.~I.~Noaki,
``Finite temperature study of the axial U(1) symmetry on the lattice with overlap fermion formulation,''
Phys. Rev. D \textbf{87} (2013) no.11, 114514
[erratum: Phys. Rev. D \textbf{88} (2013) no.1, 019901]
doi:10.1103/PhysRevD.87.114514
[arXiv:1304.6145 [hep-lat]].

\bibitem{Buchoff:2013nra} 
  M.~I.~Buchoff {\it et al.},
  ``QCD chiral transition, U(1)A symmetry and the dirac spectrum using domain wall fermions,''
  Phys.\ Rev.\ D {\bf 89}, no. 5, 054514 (2014)
  doi:10.1103/PhysRevD.89.054514
  [arXiv:1309.4149 [hep-lat]].
  
\bibitem{Dick:2015twa} 
  V.~Dick, F.~Karsch, E.~Laermann, S.~Mukherjee and S.~Sharma,
  ``Microscopic origin of $U_A(1)$ symmetry violation in the high temperature phase of QCD,''
  Phys.\ Rev.\ D {\bf 91}, no. 9, 094504 (2015)
  doi:10.1103/PhysRevD.91.094504
  [arXiv:1502.06190 [hep-lat]].
  
\bibitem{Brandt:2016daq} 
  B.~B.~Brandt, A.~Francis, H.~B.~Meyer, O.~Philipsen, D.~Robaina and H.~Wittig,
  ``On the strength of the $U_A(1)$ anomaly at the chiral phase transition in $N_f=2$ QCD,''
  JHEP {\bf 1612}, 158 (2016)
  doi:10.1007/JHEP12(2016)158
  [arXiv:1608.06882 [hep-lat]].

  
\bibitem{Ishikawa:2017nwl} 
  K.-I.~Ishikawa, Y.~Iwasaki, Y.~Nakayama and T.~Yoshie,
  ``Nature of chiral phase transition in two-flavor QCD,''
  arXiv:1706.08872 [hep-lat].


\bibitem{Bazavov:2019www} 
  A.~Bazavov {\it et al.},
  Phys.\ Rev.\ D {\bf 100}, no. 9, 094510 (2019)
  doi:10.1103/PhysRevD.100.094510
  [arXiv:1908.09552 [hep-lat]].

\bibitem{Kaczmarek:2023bxb}
O.~Kaczmarek, R.~Shanker and S.~Sharma,
Phys. Rev. D \textbf{108} (2023) no.9, 094501
doi:10.1103/PhysRevD.108.094501
[arXiv:2301.11610 [hep-lat]].


\bibitem{Neubert:1993mb}
M.~Neubert,
``Heavy quark symmetry,''
Phys. Rept. \textbf{245}, 259-396 (1994)
doi:10.1016/0370-1573(94)90091-4
[arXiv:hep-ph/9306320 [hep-ph]].

\bibitem{Brambilla:2004jw}
N.~Brambilla, A.~Pineda, J.~Soto and A.~Vairo,
``Effective Field Theories for Heavy Quarkonium,''
Rev. Mod. Phys. \textbf{77}, 1423 (2005)
doi:10.1103/RevModPhys.77.1423
[arXiv:hep-ph/0410047 [hep-ph]].

\bibitem{Laine:2003bd}
M.~Laine and M.~Vepsalainen,
``Mesonic correlation lengths in high temperature QCD,''
JHEP \textbf{02} (2004), 004
doi:10.1088/1126-6708/2004/02/004
[arXiv:hep-ph/0311268 [hep-ph]].




\bibitem{Glozman:2014mka}
L.~Y.~Glozman,
``SU(4) symmetry of the dynamical QCD string and genesis of hadron spectra,''
Eur. Phys. J. A \textbf{51}, no.3, 27 (2015)
doi:10.1140/epja/i2015-15027-x
[arXiv:1407.2798 [hep-ph]].

\bibitem{Glozman:2016swy}
L.~Y.~Glozman,
``$SU(2N_F)$ symmetry of QCD at high temperature and its implications,''
Acta Phys. Polon. Supp. \textbf{10}, 583 (2017)
doi:10.5506/APhysPolBSupp.10.583
[arXiv:1610.00275 [hep-lat]].


\bibitem{Glozman:2017dfd} 
  L.~Y.~Glozman,
  ``Chiralspin symmetry and QCD at high temperature,''
  Eur.\ Phys.\ J.\ A {\bf 54}, no. 7, 117 (2018)
  doi:10.1140/epja/i2018-12560-0
  [arXiv:1712.05168 [hep-ph]].

\bibitem{Lang:2018vuu} 
  C.~B.~Lang,
  ``Low lying eigenmodes and meson propagator symmetries,''
  Phys.\ Rev.\ D {\bf 97}, no. 11, 114510 (2018)
  doi:10.1103/PhysRevD.97.114510, 10.1103/PHYSREVD.97.114510
  [arXiv:1803.08693 [hep-ph]].


\bibitem{Catillo:2021rrq}
M.~Catillo,
``On SU(2)CS-like groups and invariance of the fermionic action in QCD,''
Int. J. Mod. Phys. A \textbf{37}, no.16, 2250102 (2022)
doi:10.1142/S0217751X22501020
[arXiv:2109.03532 [hep-lat]].

\bibitem{Glozman:2022lda}
L.~Y.~Glozman, O.~Philipsen and R.~D.~Pisarski,
``Chiral spin symmetry and the QCD phase diagram,''
Eur. Phys. J. A \textbf{58}, no.12, 247 (2022)
doi:10.1140/epja/s10050-022-00895-4
[arXiv:2204.05083 [hep-ph]].



\bibitem{Cossu:2015kfa}
G.~Cossu \textit{et al.} [JLQCD],
``Violation of chirality of the M\"obius domain-wall Dirac operator from the eigenmodes,''
Phys. Rev. D \textbf{93} (2016) no.3, 034507
doi:10.1103/PhysRevD.93.034507
[arXiv:1510.07395 [hep-lat]].


\bibitem{Tomiya:2016jwr}
A.~Tomiya, G.~Cossu, S.~Aoki, H.~Fukaya, S.~Hashimoto, T.~Kaneko and J.~Noaki,
``Evidence of effective axial U(1) symmetry restoration at high temperature QCD,''
Phys. Rev. D \textbf{96} (2017) no.3, 034509
doi:10.1103/PhysRevD.96.034509
[arXiv:1612.01908 [hep-lat]].


\bibitem{Aoki:2020noz}
S.~Aoki \textit{et al.} [JLQCD],
``Study of the axial $U(1)$ anomaly at high temperature with lattice chiral fermions,''
Phys. Rev. D \textbf{103}, no.7, 074506 (2021)
doi:10.1103/PhysRevD.103.074506
[arXiv:2011.01499 [hep-lat]].




\bibitem{Kaplan:1992bt} 
  D.~B.~Kaplan,
  ``A Method for simulating chiral fermions on the lattice,''
  Phys.\ Lett.\ B {\bf 288}, 342 (1992)
  doi:10.1016/0370-2693(92)91112-M
  [hep-lat/9206013].

\bibitem{Shamir:1993zy} 
  Y.~Shamir,
  Nucl.\ Phys.\ B {\bf 406}, 90 (1993)
  doi:10.1016/0550-3213(93)90162-I.
  [hep-lat/9303005].
  
\bibitem{Furman:1994ky} 
  V.~Furman and Y.~Shamir,
  ``Axial symmetries in lattice QCD with Kaplan fermions,''
  Nucl.\ Phys.\ B {\bf 439}, 54 (1995)
  doi:10.1016/0550-3213(95)00031-M.
  [hep-lat/9405004].



\bibitem{Brower:2005qw} 
  R.~C.~Brower, H.~Neff and K.~Orginos,
  ``Mobius fermions,''
  Nucl.\ Phys.\ Proc.\ Suppl.\  {\bf 153}, 191 (2006)
  doi:10.1016/j.nuclphysbps.2006.01.047
  [hep-lat/0511031].

\bibitem{Brower:2012vk} 
  R.~C.~Brower, H.~Neff and K.~Orginos,
  ``The Möbius domain wall fermion algorithm,''
  Comput.\ Phys.\ Commun.\  {\bf 220}, 1 (2017)
  doi:10.1016/j.cpc.2017.01.024
  [arXiv:1206.5214 [hep-lat]].

\bibitem{Neuberger:1997fp} 
  H.~Neuberger,
  ``Exactly massless quarks on the lattice,''
  Phys.\ Lett.\ B {\bf 417}, 141 (1998)
  doi:10.1016/S0370-2693(97)01368-3
  [hep-lat/9707022].

\bibitem{Aoki:2024uvl}
S.~Aoki \textit{et al.} [JLQCD:],
``Chiral susceptibility and axial U(1) anomaly near the (pseudo-)critical temperature,''
PoS \textbf{LATTICE2023} (2024), 184
doi:10.22323/1.453.0184
[arXiv:2401.06459 [hep-lat]].

\bibitem{Mathur:2006bs}
N.~Mathur, A.~Alexandru, Y.~Chen, S.~J.~Dong, T.~Draper, I.~Horvath, F.~X.~Lee, K.~F.~Liu, S.~Tamhankar and J.~B.~Zhang,
Phys. Rev. D \textbf{76} (2007), 114505
doi:10.1103/PhysRevD.76.114505
[arXiv:hep-ph/0607110 [hep-ph]].


\bibitem{Rohrhofer:2017grg} 
  C.~Rohrhofer, Y.~Aoki, G.~Cossu, H.~Fukaya, L.~Y.~Glozman, S.~Hashimoto, C.~B.~Lang and S.~Prelovsek,
  ``Approximate degeneracy of $J=1$ spatial correlators in high temperature QCD,''
  Phys.\ Rev.\ D {\bf 96}, no. 9, 094501 (2017)
  Erratum: [Phys.\ Rev.\ D {\bf 99}, no. 3, 039901 (2019)]
  doi:10.1103/PhysRevD.96.094501, 10.1103/PhysRevD.99.039901
  [arXiv:1707.01881 [hep-lat]].

\bibitem{Rohrhofer:2019qwq} 
  C.~Rohrhofer {\it et al.},
  ``Symmetries of spatial meson correlators in high temperature QCD,''
  Phys.\ Rev.\ D {\bf 100}, no. 1, 014502 (2019)
  doi:10.1103/PhysRevD.100.014502
  [arXiv:1902.03191 [hep-lat]].
  
\bibitem{Rohrhofer:2019qal}
C.~Rohrhofer, Y.~Aoki, L.~Y.~Glozman and S.~Hashimoto,
``Chiral-spin symmetry of the meson spectral function above $T_c$,''
Phys. Lett. B \textbf{802}, 135245 (2020)
doi:10.1016/j.physletb.2020.135245
[arXiv:1909.00927 [hep-lat]].


\bibitem{Chiu:2023hnm}
T.~W.~Chiu,
``Symmetries of meson correlators in high-temperature QCD with physical (u/d,s,c) domain-wall quarks,''
Phys. Rev. D \textbf{107}, no.11, 114501 (2023)
doi:10.1103/PhysRevD.107.114501
[arXiv:2302.06073 [hep-lat]].


\bibitem{Chiu:2024jyz}
T.~W.~Chiu,
``Symmetries of spatial correlators of light and heavy mesons in high temperature lattice QCD,''
Phys. Rev. D \textbf{110}, no.1, 014502 (2024)
doi:10.1103/PhysRevD.110.014502
[arXiv:2404.15932 [hep-lat]].

\bibitem{Chiu:2024bqx}
T.~W.~Chiu,
``Symmetries in high-temperature lattice QCD with (u, d, s, c, b) optimal domain-wall quarks,''
[arXiv:2411.16705 [hep-lat]].


\bibitem{Ward:2024wze}
D.~Ward, S.~Aoki, Y.~Aoki, H.~Fukaya, S.~Hashimoto, I.~Kanamori, T.~Kaneko, J.~Goswami and Y.~Zhang,
``Study of symmetries in finite temperature $N_f=2$ QCD with M\"obius Domain Wall Fermions,''
[arXiv:2412.06574 [hep-lat]].

\bibitem{Ward:2024tdm}
D.~Ward, S.~Aoki, Y.~Aoki, H.~Fukaya, S.~Hashimoto, I.~Kanamori, T.~Kaneko, J.~Goswami and Y.~Zhang,
``Study of Chiral Symmetry and $U(1)_A$ using Spatial Correlators for $N_f$ = 2 + 1 QCD at finite temperature with Domain Wall Fermions,''
PoS \textbf{LATTICE2023} (2024), 182
doi:10.22323/1.453.0182
[arXiv:2401.07514 [hep-lat]].



\bibitem{Aoki:2021qws}
S.~Aoki \textit{et al.} [JLQCD],
``Role of the axial U(1) anomaly in the chiral susceptibility of QCD at high temperature,''
PTEP \textbf{2022}, no.2, 023B05 (2022)
doi:10.1093/ptep/ptac001
[arXiv:2103.05954 [hep-lat]].




\bibitem{JLQCD:2024xey}
S.~Aoki \textit{et al.} [JLQCD],
``Axial U(1) symmetry near the pseudocritical temperature in $N_f=2+1$ lattice QCD with chiral fermions,''
PoS \textbf{LATTICE2023} (2024), 185
doi:10.22323/1.453.0185
[arXiv:2401.14022 [hep-lat]].


\bibitem{Goswami:2024kcq}
J.~Goswami \textit{et al.} [JLQCD],
``Characterizing Strongly Interacting Matter at Finite Temperature: (2+1)-Flavor QCD with M\"obius Domain Wall fermions,''
PoS \textbf{LATTICE2023} (2024), 187
doi:10.22323/1.453.0187



\bibitem{Cossu:2013ola}
G.~Cossu, J.~Noaki, S.~Hashimoto, T.~Kaneko, H.~Fukaya, P.~A.~Boyle and J.~Doi,
``JLQCD IroIro++ lattice code on BG/Q,''
[arXiv:1311.0084 [hep-lat]].


\bibitem{Boyle:2015tjk}
P.~Boyle, A.~Yamaguchi, G.~Cossu and A.~Portelli,
``Grid: A next generation data parallel C++ QCD library,''
[arXiv:1512.03487 [hep-lat]].


\bibitem{Ueda:2014rya}
S.~Ueda, S.~Aoki, T.~Aoyama, K.~Kanaya, H.~Matsufuru, S.~Motoki, Y.~Namekawa, H.~Nemura, Y.~Taniguchi and N.~Ukita,
``Development of an object oriented lattice QCD code 'Bridge++',''
J. Phys. Conf. Ser. \textbf{523}, 012046 (2014)
doi:10.1088/1742-6596/523/1/012046

\bibitem{Amagasa:2015zwb}
T.~Amagasa, S.~Aoki, Y.~Aoki, T.~Aoyama, T.~Doi, K.~Fukumura, N.~Ishii, K.~I.~Ishikawa, H.~Jitsumoto and H.~Kamano, \textit{et al.}
``Sharing lattice QCD data over a widely distributed file system,''
J. Phys. Conf. Ser. \textbf{664}, no.4, 042058 (2015)
doi:10.1088/1742-6596/664/4/042058




\bibitem{Krasniqi:2024kwm}
A.~Krasniqi, M.~C\`e, R.~J.~Hudspith and H.~B.~Meyer,
Phys. Rev. D \textbf{110} (2024) no.11, 114506
doi:10.1103/PhysRevD.110.114506
[arXiv:2407.01657 [hep-lat]].
  
\bibitem{DallaBrida:2021ddx}
M.~Dalla Brida, L.~Giusti, T.~Harris, D.~Laudicina and M.~Pepe,
``Non-perturbative thermal QCD at all temperatures: the case of mesonic screening masses,''
JHEP \textbf{04}, 034 (2022)
doi:10.1007/JHEP04(2022)034
[arXiv:2112.05427 [hep-lat]].

\bibitem{Laudicina:2022ghk}
D.~Laudicina, M.~Dalla Brida, L.~Giusti, T.~Harris and M.~Pepe,
PoS \textbf{LATTICE2022} (2023), 182
doi:10.22323/1.430.0182
[arXiv:2212.02167 [hep-lat]].

\bibitem{Giusti:2024ohu}
L.~Giusti, T.~Harris, D.~Laudicina, M.~Pepe and P.~Rescigno,
Phys. Lett. B \textbf{855} (2024), 138799
doi:10.1016/j.physletb.2024.138799
[arXiv:2405.04182 [hep-lat]].

\bibitem{Gavai:2024mcj}
R.~V.~Gavai, M.~E.~Jaensch, O.~Kaczmarek, F.~Karsch, M.~Sarkar, R.~Shanker, S.~Sharma, S.~Sharma and T.~Ueding,
``Aspects of the chiral crossover transition in (2+1)-flavor QCD with M\"obius domain-wall fermions,''
[arXiv:2411.10217 [hep-lat]].



  

  
\bibitem{Luscher:1985zq} 
  M.~Luscher and P.~Weisz,
  Phys.\ Lett.\  {\bf 158B}, 250 (1985).
  doi:10.1016/0370-2693(85)90966-9


  

\bibitem{Morningstar:2003gk}
C.~Morningstar and M.~J.~Peardon,
Phys. Rev. D \textbf{69}, 054501 (2004)
doi:10.1103/PhysRevD.69.054501
[arXiv:hep-lat/0311018 [hep-lat]].



\bibitem{Sommer:2014mea}
R.~Sommer,
``Scale setting in lattice QCD,''
PoS \textbf{LATTICE2013}, 015 (2014)
doi:10.22323/1.187.0015
[arXiv:1401.3270 [hep-lat]].

\end{thebibliography}
\end{document}